\documentstyle[times,graphics,astrobib,amssymb,epsfig,subfigure,rotating]{mn2e}

\def\labequn #1{\label{eq:#1}}
\def\labfig #1{\label{fig:#1}}
\def\labsecn #1{\label{sec:#1}}
\def\labsubsecn #1{\label{subsecn:#1}}
\def\labtablem #1{\label{tab:#1}}

\def\equn #1{Equation~\ref{eq:#1}}
\def\fig #1{Figure~\ref{fig:#1}}
\def\secn #1{Section~\ref{sec:#1}}
\def\subsecn #1{Section~\ref{subsecn:#1}}
\def\tablem #1{Table~\ref{tab:#1}}

\def\unit #1{\,{\rm #1}}
\def\nh{N_{\rm H}}
\def\cmsqi{\unit{cm^{-2}}}
\def\onlyten#1{10^{#1}}
\def\kel{\unit{K}}
\def\pc{\unit{pc}}
\def\cmcui{\unit{cm^{-3}}}
\def\ev{\unit{eV}}
\def\kev{\unit{keV}}

\def\hbeta{{\rm H\,\beta }}
\def\zsol{Z_{\odot}}

\def\etal{et al.\ }
\def\ie{i.\,e.\,, }

\title[Stability curves for Warm Absorbers in AGN]
{Properties of warm absorbers in active galaxies: a systematic stability curve analysis}
\author[Chakravorty \etal]
{Susmita Chakravorty$^{1}$
\thanks{E-mail: susmita@iucaa.ernet.in (SC); akk@iucaa.ernet.in (AK); elvis@head.cfa.harvard.edu (MA); gary@pa.uky.edu (GF)},
Ajit K. Kembhavi$^{1*}$, Martin Elvis$^{2*}$, Gary Ferland$^{3*}$ \\
$^{1}$IUCAA, Post Bag 4, Ganeshkhind, Pune 411 007, India; \\
$^{2}$Harvard-Smithsonian Center for Astrophysics, Cambridge, MA 02138; \\
$^{3}$Department of Physics and Astronomy, University of Kentucky, Lexington, KY 40506}

\begin{document}

\maketitle


\begin{abstract}

Signatures of warm absorbers are seen in soft X-ray spectra of about half of all
Seyfert1 galaxies observed and in some quasars and blazars. We use the thermal
equilibrium curve to study the influence of the shape of the ionizing continuum, 
density and the chemical composition of the absorbing gas on the existence and 
nature of the warm absorbers. We describe circumstances in which a stable warm 
absorber can exist as a multiphase medium or one with continuous variation in 
pressure. In particular we find the following results: i) the warm absorber 
exists only if the spectral index of the X-ray power-law ionizing continuum 
$\alpha > 0.2$ and has a multiphase nature if $\alpha \sim 0.8$, which 
interestingly is the spectral index for most of the observed Seyfert 1 galaxies; 
ii) thermal and ionization states of highly dense warm absorbers are 
sensitive to their density if the ionizing continuum is sufficiently soft, i.e. 
dominated by the ultraviolet iii) absorbing gas with super-Solar metallicity is 
more likely to have a multiphase nature; iv) the nature of the warm absorber is 
significantly influenced by the absence of iron and associated elements which 
are produced in the later stages of star formation history in supernovae of 
type Ia.

\end{abstract}



\section{Introduction}
\labsecn{introduction}

Absorption due to highly ionized Oxygen and elements of similar atomic number
(e.g. O~VII, ~VIII, Ne~X) is commonly found in the soft X-ray spectra of active
galactic nuclei (AGNs) (Halpern 1984; Nandra \& Pounds, 1994; Reynolds, 1997;
George $\etal$  1998).  High resolution X-ray spectra from {\em Chandra} and {\em
XMM-Newton} show that these absorption lines are always blue-shifted compared
with the emission lines and so must be in outflow from the central continuum
source (Collinge $\etal$  2001, Kaastra $\etal$  2002, Kinkhabwala $\etal$  2002, 
Blustin $\etal$  2003, Krongold
$\etal$  2003, Netzer $\etal$  2003, Turner $\etal$  2004). The absorption edges and
lines respond to variation of the ionizing continuum emitted by the central
engine. For example, Krongold et al. (2007) show that for NGC4051 the ionization 
state of the WA responds to the continuum flux as predicted for gas that stays close 
to photoionization equilibrium, 
so that other factors, e.g. velocity changes and shocks, do not appear to be
important. It is therefore reasonable to assume that the absorbing cloud is
photoionized by the continuum radiation from the active galactic nuclei
(AGN). This partially ionized optically thin gas along the line of sight to the
center of the AGN has come to be known as the warm absorber (hereafter WA).

Understanding the structure of the WA will lead to better understanding of the
structure of AGNs; for example, identifying the location from which the 
wind arises will impose constraints on the physics of the wind source. Some 
authors find that the WA is well fitted with a model with two ionization
phases (Krongold $\etal$  2003; Krongold $\etal$  2005a; Ashton $\etal$  2006; 
Collinge
$\etal$  2001) or three phases (Netzer $\etal$  2003; Morales, Fabian \& Reynolds,
2000; Kaastra $\etal$  2002), in which the components are in pressure equilibrium.
Other authors (Ogle $\etal$, 2004) and Steenbrugge $\etal$, (2005) have argued in
favor of a continuous distribution of ionization state for the WA, in objects
for which others fit just a few phases. One way to investigate which solution is
more likely is to investigate the physics determining the thermal structure of WA.

The effect of various factors on the WA can be studied conveniently using the thermal 
equilibrium curve (hereafter referred to as the {\it stability curve}) of temperature 
(T) against the ratio of ionization parameter ($\xi$, see below for definition) to T. 
Gas lying off the stability curve will cool or heat until reaching the curve. If the 
curve has kinks that produce multiple stable values at fixed $\xi/T$, then the WA can 
have multiple phases in mutual equilibrium. This effect may be responsible for the 
multiphase nature of the interstellar medium of galaxies (Field, 1965). Ionizing 
continuum from AGN can 
induce such kinks (Krolik, McKee \& Tarter, 1981), to a larger extent than the 
interstellar medium. The pronounced {\it kinks} representing WA in the stability 
curve lie in the intermediate temperature range, $5 < \log T < 7$ where photoionization 
heating and collisional cooling of various metals are expected to determine the thermal 
balance of the WA. The details of the thermal stability is 
strongly affected by the shape of the ionizing continuum emitted by the central 
source (i.e. their spectral energy distributions or SEDs), and by the chemical 
composition of the absorbing gas.

Our main motivation in this paper is to improve on previous work (Hess, Kahn \& 
Paerels 1997; Komossa \& Mathur 2001; Krolik, McKee \& Tarter 1981, Krolik \&; Kriss 2001; 
Reynolds \& Fabian 1995; Rozanska, Kowalska \& Goncalves 2008) by carrying out a systematic investigation of the 
stability curves and finally to give a quantitative handle on the relative stability 
nature of the WA as a function of the various physical parameters involved. In 
\secn{warmabs} we discuss the physical properties of the WA. In \secn{constraints} we 
give all the constraints and assumptions used in the paper. The simulations of the 
stability curves are sensitively dependent on the underlying atomic physics used by the 
photoionization codes. We have shown (Chakravorty $\etal$, 2008, hereafter Paper I), that 
dielectronic recombination rate coefficients have a large effect on the stability curve. 
Using recent atomic physics databases that include many more of these rates, with more 
accurate values, is important because it significantly changes the predictions for the 
physical nature of the WA. The serious impact of this effect is addressed in further 
details in \secn{stability} while we discuss the stability curve and comparison with 
earlier works. We choose a specific scheme to represent the ionizing continuum (discussed 
in \secn{constraints}) and vary its parameters to create a reasonable ensemble of spectral 
shapes as demonstrated by a wide variety of AGN which harbour WA. The effect of these 
different SEDs on WA nature is discussed in \secn{influencespectral}. Rozanska \etal (2008) have 
demonstrated that for sufficiently soft AGN SED, the thermal balance becomes dependent on 
the density values if it is above a certain threshold. This can only happen if 
the gas is dense enough and the ionizing continuum is appropriate to induce heating 
processes other than photoionization to dominate the thermal balance. Since we sample 
a rather large variety of ionizing continua for our investigations, the present work 
gives us an 
appropriate opportunity to study the effect of density on stability curves in 
\secn{density}. Optical and ultraviolet emission and absorption line studies suggest that 
central regions of AGN have solar or higher metallicities (Hamann \& Ferland, 1999) and 
the WA temperature range of the stability curve is sensitive to the atomic interactions 
due to various metals. This abundance dependence provides the motivation for our study of 
the influence of individual elements, and groups of elements in \secn{chemicalcomp}. The 
improved updated atomic database used for the simulations provides the opportunity to go 
beyond the qualitative discussions and, for the first time, make quantitative comparisons 
of the multiphase nature and the robustness of the stability regions as a function of the 
ionizing SED and chemical composition of the absorber in \secn{multiphase}. Our results 
are summarised in \secn{summary}.


\section{Warm Absorbers}
\labsecn{warmabs}

Studies of WA variability in response to continuum changes show that it is 
reasonable to assume the WA to be in ionization and thermal equilibrium 
(Nicastro $\etal$, 1999). The timescales to reach equilibrium seem to be less than 
a day (Nicastro $\etal$  1999; Krongold $\etal$  2003; Krongold $\etal$  2005a; 
Krongold $\etal$  2005b; Krongold $\etal$  2007). 
Equilibrium is reached when photoionization is balanced by
recombination, and ionization and Compton heating are balanced by recombination 
and Compton cooling.
cooling. The Compton heating and cooling become important at high temperature 
which is solely determined by the shape of the illuminating continuum, and typically 
is $T \gtrsim 10^7~K$ for standard Seyfert 1 spectral energy distributions, 
which the clouds can attain only when the ionizing continuum 
has significant number of hard X-ray photons. The typical
column density in WAs, estimated mainly using absorption edges, is
$\nh\sim\onlyten{22\pm1}\cmsqi$ (see the references in \secn{introduction}). 
In some cases lines are easily detected whereas edges are not seen 
(Kaastra $\etal$  2002; Rozanska $\etal$  2004). However, the $\nh$ values derived 
from edges alone are systematically high by large factors ($\sim$10) - see 
Krongold $\etal$  2003, Kaspi $\etal$  2002.
Modeling of the absorber allows the temperature and ionization parameter to be
calculated. Typically, the temperatures are of the order of $\onlyten{5}\kel$
and the ionization parameter $\xi\sim100$ (see definition of $\xi$ below).
However, in some objects WAs are found to have additional components which exhibit 
higher temperatures, $\sim 10^6 \kel$ (see the references in \secn{introduction}).

The estimated distance of the WA from the central source covers a wide range:
from the kiloparsec scale of the extended narrow emission line region
(Kinkhabwala $\etal$  2002), through distances of $\sim0.01-1\pc$, \ie at the scale
of the obscuring molecular torus (Krolik \& Kriss, 2001; Crenshaw $\etal$  2003;
Blustin $\etal$  2005), to within the $\hbeta$ broad emission line region, on the
scale of the accretion disk, which is a few thousand Schwartzchild radii 
(Krongold $\etal$, 2007), where the WA would form the base of an accretion disk wind
(Murray \& Chiang 1995, Elvis 2000). The photoionization state of the WA can 
be parametrised by the ionization parameter which is the ratio of the ionizing photon 
flux to the gas density (Tarter, Tucker \& Salpeter, 1969):
\begin{equation}
\xi = L_{\rm{ion}}/n_{\rm{H}}\,R^2 ~~~~~~~\unit{[erg\,cm\,s^{-1}]},
\labequn{ionparm}
\end{equation}
where $L_{\rm{ion}}$ is the luminosity between $1 - 10^3$ Rydberg.
Hence $\xi/T \sim L/pR^2$, $p$ being the gas pressure in the WA. Although, 
in this paper we have used $\xi$ for all our discussions, in the context of warm 
absorbers, there is another popular definition of the ionization parameter given by
\begin{equation}
U = Q_{\rm{ion}}/4\,\pi\,c\,n_{\rm{H}}\,R^2.
\labequn{U}
\end{equation}
$Q_{\rm{ion}}$ ($\rm{s^{-1}}$) is the rate of incident photons above 1 
Rydberg and c is the velocity of light. For a given shape of the SED 
of the ionizing continuum, the conversion from $\xi$ to 
$U$ is unique and can be done conveniently. Although luminosity can be 
constrained from separate independent observations, it is difficult to remove the 
degeneracy between density and distance in the product $n_{\rm{H}}\,R^2$ featuring in 
the denominator of the ionization parameter. Thus, the estimation of distances for WA 
becomes challenging; for example, Netzer \etal 2003 and Krongold \etal 2006 put upper 
limits differing by up to three orders of magnitude on the WA 
distance in the same object NGC3783 (but using different 
observations). As the mass loss rate is proportional
to the radius such large factors of uncertainty have a large effect on AGN 
energetics and on feedback from the AGN to the surrounding media.

Krolik, McKee and Tarter (1981) showed that a stable state exists at
$T\sim10^4\kel$ for a photoionized cloud which undergoes collisional cooling.
There can be an additional stable state for the absorbing cloud at
$T\sim10^8\kel$, if the incident photoionizing continuum has a hard X-ray
component as an extra heating agent. Thus they suggest that presence of a hard
X-ray continuum would result in clumping, with cold clouds remaining in
equilibrium with a hot medium. However, for optically thin, ionized gas, the
temperature range $\onlyten{5}-\onlyten{7}\kel$ is well known to be an unstable
zone, bringing into question the model of WA as gas in thermal
equilibrium. Radiation in spectral lines is the primary cooling mechanism in
this temperature range. The contribution from radiative recombination and
two-photon continuum emission (Gehrels \& Williams 1993) form some additional
small regions of stability.

\section{Constraints and Assumptions}
\labsecn{constraints}

\subsection{The code and its implementation}
\labsubsecn{constraintsCode}

We assume a simple model of a geometrically thick, optically thin, plane
parallel slab of gas illuminated on one face for the absorbing medium which is
located at a certain distance $R$ from a central continuum source of bolometric 
luminosity $L$. The geometrical thickness of the gas is taken to be much less 
than $R$. The physical condition of the absorbing medium is determined by the 
flux received at the illuminated face of the gas ($L/4\pi R^2$), the shape of 
the ionizing continuum (SED), the number density of the gas $n_{\rm{H}}$, the 
column density of the 
gas $\nh$, and the chemical composition of the absorbing gas. To obtain the
equilibrium conditions for a gas with an assumed set of parameters which specify
these inputs, we have used the publicly available photoionization code 
{\tiny CLOUDY}\footnotemark version C07.02 (hereafter C07), see Ferland (1998).
\footnotetext{URL: http://www.nublado.org/ } 

{\tiny CLOUDY} calculates the ionization equilibrium conditions by solving the energy
and charge conservation equations under the assumption that all the atomic
processes have had time to reach a steady state. The density $n_i$ of $i^{th}$
species is then given by the balance equation
\begin{eqnarray}
\frac{\partial n_i}{\partial t} & = & \sum \limits_{j \ne i} n_j R_{ji} + Source\\ \nonumber
 & & ~~~~~~ - n_i \left( \sum \limits_{j \ne i} R_{ij} + Sink \right) \\ \nonumber
& = & 0,
\labequn{iondensity} 
\end{eqnarray}
where $R_{ji}$ is the rate of species $j$ going to $i$, {\it Source} is the 
rate per unit volume of appearance of new atoms in $i^{th}$ species and 
{\it Sink} is the rate that they are lost (Osterbrock \& Ferland, 2006). 
Thermal equilibrium conditions are calculated by evaluating the heating and
cooling rates at different temperatures for the given gas and finding the
temperatures where the two rates balance each other.

For all the calculations in this paper we have used {\tiny CLOUDY} in the 
{\it constant density} mode which is chosen for the sake of numerical 
simplicity, as we probe a rather large range of parameter space to 
determine the properties of the WA. We would like to note here that using the 
TITAN code (Dumont, Abrassart \& Collin, 2000), Rozanska $\etal$ (2006) and 
Goncalves 
$\etal$ (2006) have pioneered the {\it constant total pressure} mode of 
calculations for photoionization equilibrium where they relax the 
approximation that the density is constant. Their only assumption is that the line of 
sight gas is in total pressure equilibrium and they introduce additional iterations 
between temperature and density within the code in a self consistent manner. 
This results in natural 
stratification of the cloud with different temperatures and densities where each 
layer is illuminated by the radiation already absorbed by the previous layer. 
These two methods agree well with each other for optically thin clouds, but 
show significant differences when the column density of the absorbing cloud is 
high ($\nh \gtrsim 10^{22}$). In our future publications where we will restrict 
ourselves to study equilibrium properties of WA as functions of a smaller range 
of physical parameters, we will use similar methods.


\subsection{Density dependence}
\labsubsecn{density}

Estimation of temperature is independent of $n_{\rm{H}}$, for a gas in thermal and 
ionization equilibrium when the dominant processes of heating and cooling are 
photoionization and recombination respectively. This is because in this regime of 
thermal equilibrium, density enters the temperature estimation only through the 
ionization parameter. Hence the stability curve would be insensitive to the value of 
the density used. This gives us the liberty to choose an intermediate value for WA 
density, $n_{\rm{H}} \sim\onlyten{9}\cmcui$, leading to $L/R^2\sim\onlyten{11}$ from
\equn{ionparm}, for typical values $\xi\sim100 $ corresponding to WA. Except for 
\secn{density}, where we check the robustness of this assumption, we have assumed 
the value $\onlyten{11}$ for $L/R^2$ throughout the paper.


\subsection{Ionizing continuum}
\labsubsecn{sed}

The $1-10 \kev$ part of the observed spectra of AGN is usually modeled as a
simple power-law with exponential lower and upper cut-offs specified by
$\nu_{min}$ and $\nu_{max}$ respectively,
\begin{equation}
f(\nu) \sim\nu^{-\alpha} e^{-\frac{\nu_{min}}{\nu}} e^{-\frac{\nu}{\nu_{max}}}, 
\labequn{power-law}
\end{equation}
where $f(\nu)$ is the flux per unit frequency interval and $\alpha(>0)$ is the
energy spectral index. For most quasars, the X-ray spectral index lies in the
range $0.7<\alpha<0.9$ as shown by Wilkes \&
Elvis in 1987 and also recently by Grupe $\etal$  2006 and Lopez $\etal$  2006. 
However some objects show extreme values like
$\alpha<0$ or $\alpha>2$. There is a systematic difference between
radio-quiet and radio-loud AGNs in that radio-quite AGNs have steeper X-ray
spectra ($\alpha \sim 1.0$) than radio-loud ones, ($\alpha\sim 0.5$, Wilkes \&
Elvis 1987; Grupe $\etal$  2006; Lopez $\etal$  2006). 

Ionizing continua given by \equn{power-law} have been used earlier by 
Reynolds \& Fabian (1995, hereafter RF95) for doing detailed analysis of nature 
of WA as a function of various relevant physical parameters. Their work is further 
discussed in \secn{stability}. The dashed and dotted curve labeled as 
{\it Early set} in the lower panel of \fig{BasicCont} is drawn using \equn{power-law} 
with $\alpha = 0.8$, $E_{min} = 13.6 \ev$ and $E_{max} = 40 \kev$.

The extension of the soft X-ray SED into the extreme ultraviolet (EUV) cannot be
directly observed in most cases because energy in EUV is easily absorbed by our
own Galaxy. The join between the EUV and X-ray flux is parametrized by the slope
$\alpha_{ox}$ defined as
\begin{eqnarray}
\alpha_{ox} & = & -0.384 \log\left[\frac{F_{\nu}(2 keV)}{F_{\nu}(2500 \AA)}\right] 
\labequn{alphaOX}
\end{eqnarray}
(Tananbaum $\etal$, 1979). To obtain $\alpha_{ox}$ in the observed range 
$1 \le \alpha_{ox} \le 2$ we need a steeper soft component in addition to the 
power-law given by \equn{power-law}. We assume a two component SED to 
simultaneously include the information on X-ray and EUV and which can be 
expressed as
\begin{equation}
f(\nu) \sim \left [ \nu^{-\alpha} + \eta \nu^{-\alpha_s} \right ] e^{-\frac{\nu}{\nu_{max}}} ~~~ \rm{for} \,\, \nu \ge 2500 \AA, 
\labequn{2power-law}
\end{equation}
where $\alpha_s (> 0)$ is the spectral index of the steep soft component and 
$\eta$ is the relative normalization factor. For energies lower than $2500 \AA$ 
we have used a spectral shape given by
\begin{eqnarray}
f(\nu) & \sim & \nu^{-0.5} ~~~ \rm{for} \,\ 0.206 \,\,\rm{Ryd} \,\, \le \nu < 2500 \AA  \nonumber \\
& \sim & \nu^{-1.0} ~~~ \rm{for} \,\ 9.12 \times 10^{-3} \le \nu < 0.206 \,\,\rm{Ryd} \nonumber \\
& \sim & \nu^{2.5} ~~~ \rm{for} \,\ \nu < 9.12 \times 10^{-3} 
\labequn{Emin}
\end{eqnarray}
which is similar to the shape used by Mathew \& Ferland (1987) for the same 
energies. \equn{2power-law} and \equn{Emin} taken together describe a more 
realistic ionizing continuum which irradiates the WA. Such a continuum is shown 
as the solid curve labeled {\it Standard set} in \fig{BasicCont} with 
$\alpha = 0.8$, $\alpha_s = 2.0$, $\alpha_{ox} = 1.2$, $E_{max} = 200 \kev$ and 
lower energy cut-off following \equn{Emin}. 

\begin{figure}
\begin{center}
\includegraphics[scale = 1, width = 9 cm, trim = 150 370 150 50, clip]{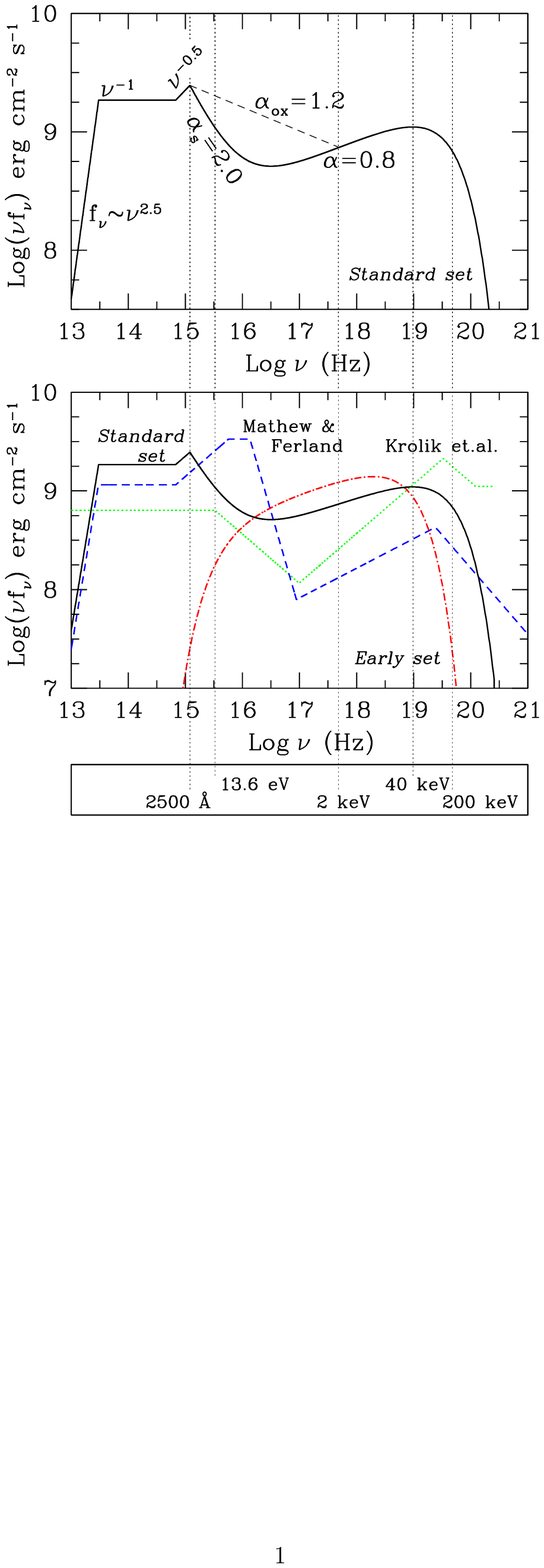}
\caption{Top panel : Ionizing continuum obtained using \equn{2power-law} and 
\equn{Emin} taken together, where $\alpha = 0.8$, $\alpha_s = 2.0$, $\alpha_{ox} = 1.2$ 
and the continuum is exponentially cut-off at $200\kev$. At lower energies, this 
continuum is cut-off following \equn{Emin} with the successive spectral indices of 
$-0.5, \,\, -1.0$ and $2.5$. This continuum is labeled as the {\it Standard set}. 
We have also appropriately labeled the dominant component of the SED at various 
energies and used the dashed line to show the resultant $\alpha_{ox} = 1.2$. 
Bottom panel : Comparison of the {\it Standard set} ionizing continuum with 
various other SEDs used or discussed in this paper. 
The dotted-and-dashed line labeled as the {\it Early set} is the ionizing continuum 
given by \equn{power-law} with a X-ray slope $\alpha = 0.8$ and exponential cut-offs 
at $E_{min} = 13.6 \ev$ and $E_{max} = 40 \kev$. 
The dotted line represents the ionizing 
continuum used by Krolik $\etal$ 1981 and the dashed line that used by 
Mathew \& Ferland (1987). The important energy values including the upper and lower 
energy cut-offs ($13.6 \ev$, $40 \kev$ and $200 \kev$) and the range of definition 
for $\alpha_{ox}$ ($2500 \AA$ and $2 \kev$) have 
been marked and labeled.}  
\labfig{BasicCont}
\end{center}
\end{figure}

In the lower panel of \fig{BasicCont}, 
the {\it Standard set} continuum is compared with various other SEDs used or 
discussed in this paper and we will refer to these comparisons in more details in 
\secn{stability}.    
In this paper we have used the simple definition for the ionizing continuum, as
in \equn{power-law}, only in \secn{stability} where we have given an introductory
description of the stability curve and compared results with similar works in
literature (RF95). In all other sections we have considered the more realistic 
{\it Standard set} spectra. All through the paper we have 
assumed $\alpha_s = 2$, and for a given value of $\alpha$, obtained the desired 
value of
$\alpha_{ox}$ by adjusting $\eta$. In \secn{influencespectral}, when we
investigate the changes in the nature of the stability curve with
variation in the shape of the continuum, we consider a wide range of slopes,
$0.2<\alpha<1.1$ and $1<\alpha_{ox}<1.8$ and a wide range of higher energy 
cutoff, $50 \kev < E_{max} < 400 \kev$ to incorporate various types of AGN.


\section{Stability Curve and earlier studies on it}
\labsecn{stability}


\subsection{Definition}
\labsubsecn{definition}

The stability curve, may be considered to be a phase
diagram in the $\log T - \log (\xi/T)$ plane, with each point on the curve
representing an equilibrium state of the system at a temperature $T$, for
ionization parameter $\xi$. The other parameters describing the system have
specific values as mentioned below. The solid curve in \fig{stability} is
generated using version C07 of {\tiny CLOUDY}, for an absorber having gas density
$n_{\rm{H}}=\onlyten{11}/\xi$ and solar metallicity, being ionized by a 
power-law continuum having energy spectral index $\alpha=0.8$ and extending 
from  $13.6 \ev \,\, \rm{to} \,\, 40 \kev$; hereafter we will refer to this 
as the {\it Early model} and the corresponding parameter set as the 
{\it Early set}. Computations were done using a grid of models with a range 
of values of the ionization parameter $\xi$,
and the code is constrained to perform single zone calculations to ensure that
the gas is optically thin. The ionizing 
continuum used in the {\it Early model} is shown as the dotted-and-dashed curve 
in  the lower panel of \fig{BasicCont}.  We use the {\it Early set} in 
this section to compare results with earlier works like RF95 who have done 
extensive study of the stability properties of the WA as a function of various 
shapes of the ionizing continuum. 

An isobaric perturbation of a system in equilibrium is represented by a small
vertical displacement from the stability curve; such perturbations leave 
$\xi/T\sim
L/pR^2$ constant, which for constant $L/R^2$ leave the pressure unchanged. If
the system is located on a part of the curve with positive slope, then a
perturbation corresponding to an increase in temperature leads to cooling, while
a decrease in temperature leads to heating of the gas. A gas in such a state is
therefore thermally stable. On the other hand, a gas located in a region of the
curve with negative slope is thermally unstable, and may achieve a multiphase
equilibrium under certain physical conditions which we will discuss in later
sections.


\subsection{Earlier studies}
\labsubsecn{earlierStudies}

Before we go on to discuss our own results with the stability curve, here, 
we would like to mention some of the earlier works done by various authors on 
WA and/or stability curves. 

Krolik $\etal$  (1981) obtained the stability curve for AGN using parameters
similar to our {\it Early set} except for the ionizing continuum. The 
continuum used by them is plotted as the dotted curve in \fig{BasicCont}. 
They have argued in favour of $\sim 10^4 \kel$ stable states remaining in 
pressure equilibrium with the $\sim 10^8 \kel$ phases. Although there are 
some {\it narrow} stable phases at $\sim 10^5 \kel$, they have ignored the 
significance of such states. Moreover their stability curves do not show any 
stable states at $\sim 10^6 \kel$. 

In a similar study on the absorbing medium,
Mathew \& Ferland (1987) do not find 
the $\sim 10^4 \kel$ and $\sim 10^8 \kel$ phases in pressure equilibrium with 
each other. The continuum used by Mathew \& Ferland (1987) is widely used as a 
broad band continuum for AGN modeling studies and has been represented with the 
dashed line in \fig{BasicCont}. The energy spectral index of the ionizing 
continuum in the {\it Standard} set is very similar to that of the 
Mathew \& Ferland SED in X-rays (0.3 - 100 $\kev$). However, it is not 
objective to make a comparison at this stage 
because, although the ionizing continuum in the {\it Standard} set given by 
\equn{2power-law} represents the overall shape of the SED 
between 3.6 $\ev$  to 2 $\kev$ using the definition of $\alpha_{ox}$, 
it is unable to give information on any underlying specific features in this 
energy range. In this paper we will constrain ourselves to use ionizing 
continuum given by \equn{2power-law} and \equn{Emin}, but will investigate 
the details of the 
spectral shape in the energy range (1 - 1000 $\ev$) in subsequent publications 
(Chakravorty $\etal$, in preparation) by incorporating the effect of soft excess 
at $\sim 150 \ev$ and a disk blackbody component at $\sim 20 \ev$.

Reynolds and Fabian (1995, RF95) have studied the warm absorber stability 
conditions using physical conditions as described by our {\it Early set} of 
parameters. They used a simple power-law for the ionizing continuum and varied 
the power-law index $\alpha$ to check for its effects on the WA and also examined 
some basic effects of adding a soft excess component to the 
power-law ionizing continuum. Using power-law continuum with $\alpha = 0.8$, they 
find that the stability curve is independent of density variation in the range 
$10 < n_{\rm{H}} < 10^{11} \,\, \rm{cm^{-3}}$

Hess \etal (1997) address the discrepancy that the physical properties inferred 
from observations of some low-mass X-ray binaries and Seyfert galaxies correspond 
to thermally unstable regions of the stability curve predicted by photoionization 
codes. They have carried out an extensive study of the stability curve as a 
function of 
various ionizing continuum and different abundance of the absorbing gas. They 
have studied the cause of the instabilities in great detail and have cautioned 
against the outdated fits and parameters used by the simulation code and also 
against the accuracy of the atomic data then used for determining the thermal 
balance.

Komossa \& Meerschweinchen (2000) and Komossa \& Mathur (2001) have extensively 
studied the stability curve respectively  
as a function of ionizing continuum and chemical composition of the absorber with the 
aim of explaining the WA in various narrow line Seyfert 1 galaxies. They find 
that steeper ionizing continuum stabilizes WA and super-Solar metallicity of the gas 
enhances the possibility of multiphase WA. 
By modeling the observations of NGC4051, indications are found that the WA in this 
object has super-Solar abundance.

\begin{figure}
\begin{center}
\includegraphics[scale = 1, width = 8.0 cm]{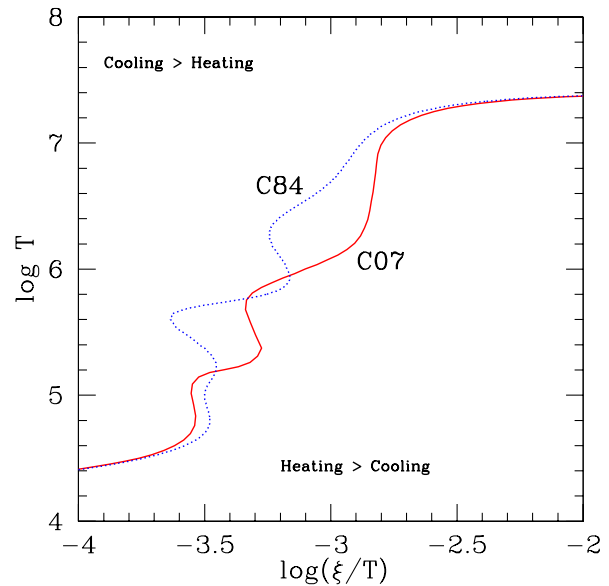}
\caption{Stability curves showing the distribution of equilibrium temperature
$T$ as a function of $\xi/T$. It is assumed that an optically thin shell of gas
of solar metallicity is ionized by a continuum with energy $\alpha = 0.8$,
extending from $13.6\ev$ to $40\kev$.  The regions of the plane where heating or
cooling dominates are indicated. The solid curve is obtained using version
C07 of {\tiny CLOUDY} while the dotted curve is for version c8412a. The {\it Early
set} of parameters, as described in the text, is used in generating the two
curves, which are seen to be significantly different for $10^{4.6} < \log T <
10^7 \kel$.}  
\labfig{stability}
\end{center}
\end{figure}


\subsection{Comparison with earlier work: importance of updated atomic data}
\labsubsecn{compare}

In the WA temperature range 
$10^5 < \log T < 10^7 \kel$ cooling processes are dominated by collisional 
recombination
which can be radiative or dielectronic. These processes lead to local regions of
thermal stability in this otherwise unstable temperature range as discussed by
Gehrels \& Williams (1993), Hess, Kahn \& Paerels (1997) and also depicted by 
RF95. We can see these 
effects in our stability curve generated using the {\it Early set} of parameter 
values and shown as the solid curve in \fig{stability}. However the atomic 
physics data base has been updated considerably over the last decade and our 
results show significant quantitative variations from earlier work 
which we emphasise by making explicit comparisons with RF95. 

The dotted 
line in \fig{stability} is a reproduction of the stability curve (using version 
C8412a of {\tiny CLOUDY}) in Figure 3 of RF95 where they had used the {\it Early set} of 
physical parameters. The variations in the C07 stability 
curve from the C84 curve lead to different physical predictions for the WA; 
for example, 
the $10^5 \kel$ state becomes much more stable in our work resulting in more 
pronounced possibility of multiphase nature of the WA. The $5 < \log T <6.5$ region 
of the stability curve, relevant for WA, sensitively depends on the detailed atomic 
physics of the various elements which contribute to photoelectric heating  and 
cooling due to recombination
and collisionally excited lines. In Paper I we have shown 
that the
total recombination rates (dielectronic + radiative), for the significant cooling 
agents which bring about the maximum variation in the stability curves in 
\fig{stability}, are predicted to be larger in C07 than in C84. 


\begin{table*}
\begin{center}
\begin{tabular}{c c c c c c c c c c c c c c c c c c c}
\hline
& & \multicolumn{11}{c}{Ionizing continuum} & & & & & &\\ \cline{3-13}
\raisebox{1.5ex}[0cm][0cm]{Model} & & & & & & & & & & & & %
\raisebox{-0.5ex}[0cm][0cm]{Low energy} & & \raisebox{1.5ex}[0cm][0cm]{$n_{\rm{H}}$} & %
& \raisebox{1.5ex}[0cm][0cm]{$N_{\rm{H}}$} & & \raisebox{1.5ex}[0cm][0cm]{Metallicity} \\
& & Definition & & $\alpha$ & & %
$\alpha_{ox}$ & & $\alpha_s$ & & $E_{max}$ & & %
\raisebox{-0.5ex}[0cm][0cm]{Cut Off} & & & & & & \\ \\
\hline \\ \\
{\it Early} & & \equn{power-law} & & 0.8 & & - & & - & & $40 \kev$ & & $E_{min} = 13.6 \ev$ & %
& $10^9 \cmcui$ & & Zone 1 & & $\zsol$ \\ 
& & & & & & & & & & & & (\equn{power-law}) & & & & (See text) & & \\ \\ 
\hline \\ \\
{\it Standard} & & \equn{2power-law} & & 0.8 & & 1.2 & & 2.0 & & $200 \kev$  & & \equn{Emin} & %
& $10^{11}/\xi$ & & $10^{22} \cmsqi$ & & $\zsol$ \\ 
& & \equn{Emin}& & & & & & & & & & & & (See text) & & & & \\ \\
\hline
\end{tabular}
\caption{Comparison of the various parameters used to constitute the {\it Early} 
and the {\it Standard} model for stability curves.}
\labtablem{table1} 
\end{center}
\end{table*}


In the WA 
temperature range $10^5 \lesssim T \lesssim
10^7\kel$, dielectronic recombination dominates over radiative
recombination for many ions (Osterbrock \& Ferland, 2006). Unlike
the radiative recombination rate coefficients, dielectronic recombination 
rate coefficients (DRRC) have
undergone significant changes over the last decade. The 
importance of revisiting the values of DRRC were first pointed out by 
Savin \etal (1997) and further emphasised by a series of papers by Badnell 
and coworkers (Colgan, Pindzola \& Badnell 2004, Colgan \etal 2003, Altun \etal 2004, 
Zatsarinny \etal 2004a, Mitnik \& Badnell 2004, Zatsarinny \etal 2003,
Gu 2003, Zatsarinny \etal 2004b and Gu 2004). The DRRC data base is 
more extensive now and shows that the rate coefficients for different 
ionization states of various elements are substantially larger than the values 
used previously, but the database is still not complete, 
especially for the lower
ionization states. For ions which do not have computed DRRC
values, C07 uses a solution, as suggested by Ali \etal (1991);
for any given kinetic temperature, ions that lack data are given
DRRC values that are the averages of all ions with the same
charge. The advantage of this method is that the assumed rates
are within the range of existing published rates at the given
kinetic temperature and hence cannot be drastically wrong. However, in Paper I, 
we have checked whether the points in the C07 stability curves
have computed DRRC values or guessed average values,
concentrating on the parts of the curve which have multiphase
solutions for the warm absorber ($10^5 \lesssim T \lesssim
10^6\kel$) and are different from the C84 stability curve. We
find that all the ions which act as major cooling agents for each
of these points in the stability curve have reliable computed
DRRC values which are not likely to change in near future. Thus, 
the new data base provides a more robust
measurement of the various physical parameters involved in
studying the thermal and ionization equilibrium of photoionised
gas. This not only gives us the motivation to improve upon previous qualitative 
stability curve analysis by various authors (Hess, Kahn \& 
Paerels 1997; Komossa \& Mathur 2001; Krolik, McKee \& Tarter 1981; 
Krolik \&; Kriss 2001; 
Reynolds \& Fabian 1995; Rozanska \etal 2008) but also provides the chance to extend 
their work with quantitative estimates on the nature of the multiphase WA 
in \secn{multiphase}.


\subsection{Standard set}
\labsubsecn{standardSet}

The ionizing continuum used in the {\it Early set} has now been shown by 
improved spectral UV and X-ray measurements not to correspond to observed 
continua (Zheng $\etal$  1997, Shang $\etal$  2005). As discussed in 
\secn{constraints}, we will use a more realistic SED with two power-law 
components given by \equn{2power-law} and lower energy cut-off given 
by \equn{Emin}. The fiducial value for $\alpha_{ox}$ for Seyfert1 galaxies 
is 1.2 (Netzer 1993). We define a {\it Standard set} 
of parameter values to be used in the rest of the paper where the ionizing 
continuum shown as the solid line in \fig{BasicCont} is parametrized 
by $\alpha=0.8$, $\alpha_s=2$ and $\eta$ is appropriately adjusted to 
attain $\alpha_{ox}=1.2$. $E_{max}$ for both components is $200\kev$. 
The high energy cut-off is an average of the observational values reported from 
Beppo-Sax which found hard X-ray spectral cut-off to span the range 50 
to 450 $\kev$ (Matt 2004; Tueller \etal 2008). The absorbing gas in this {\it Standard} model 
is assumed to have solar abundance ($Z=\zsol$) as given by Allende Prieto, Lambert \& Asplund 
(2002, 2001), Holweger (2001) and Grevesse \& Sauval (1998). The 
density is given by $n_{\rm{H}} =\onlyten{11}/\xi$ as in the 
{\it Early set}. In addition we assume $\nh = 10^{22} \cmsqi$, for the slab 
of the gas in the {\it Standard set}. \tablem{table1} shows a 
comparison of the various parameters used in the {\it Early set} and the 
{\it Standard set}.

\begin{figure*}
\begin{center}
\includegraphics[scale = 1, width = 19 cm, trim = 10 290 5 50, clip]{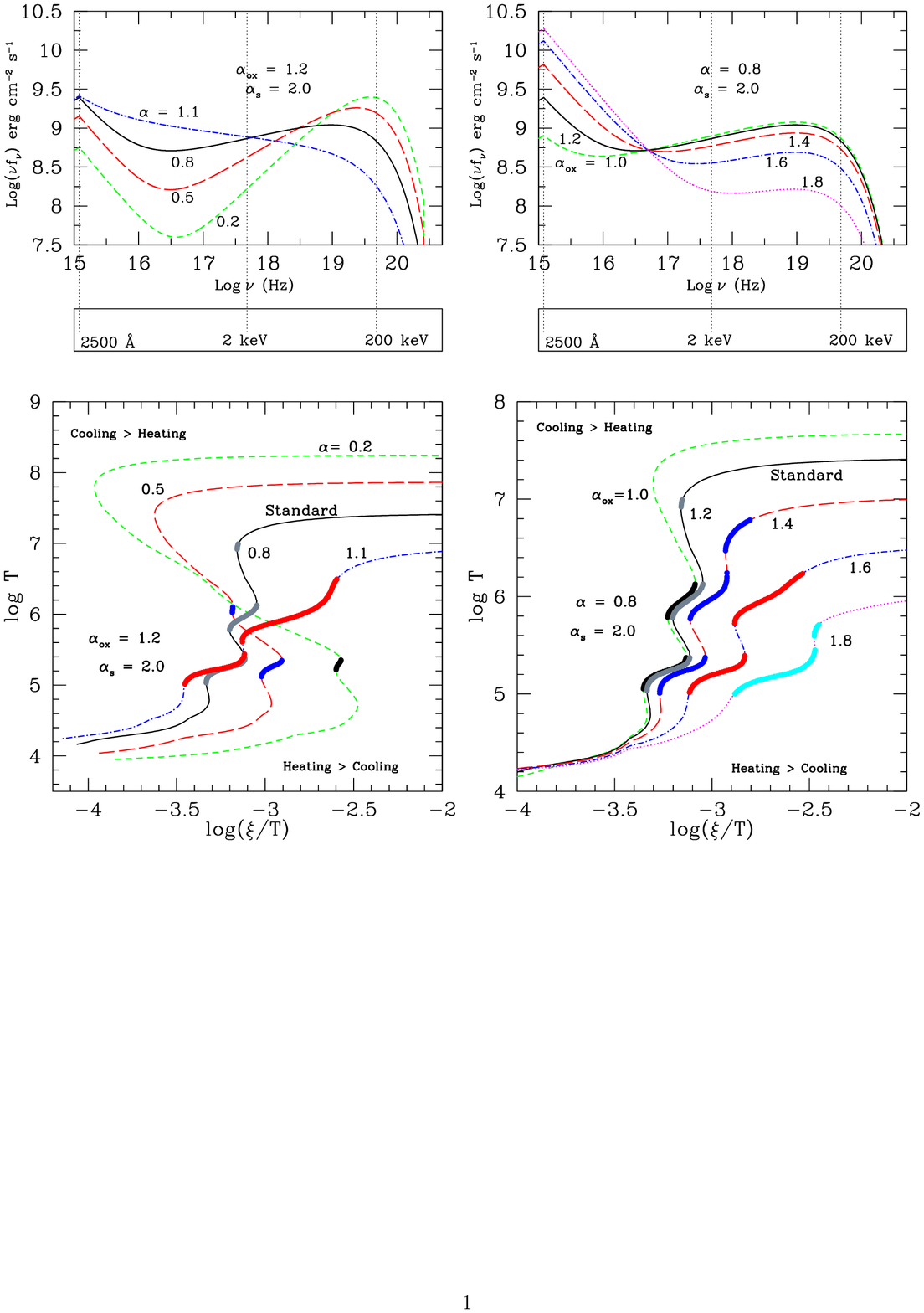}
\caption{The shapes of the ionizing continua and the corresponding 
stability curves  
for varying X-ray slope $\alpha$ (left) and varying EUV to X-ray slope, 
$\alpha_{ox}$ (right) for an optically thin shell of gas of solar 
metallicity. In the top panels showing the SEDs, the important energy 
values including the upper energy cut-off (200 \kev) 
and the range of definition for $\alpha_{ox}$ ($2500 \AA$ and 2 \kev) have 
been marked and labeled. Left panels: X-ray flux index $\alpha$ is varied 
from one curve to the other in the range $0.2<\alpha<1.1$ with $\alpha_{ox}$ 
held constant at the value $1.2$. Right panels: The EUV to soft X-ray slope 
$\alpha_{ox}$ is varied over a range of values with the X-ray slope $\alpha$ 
held constant at the standard value $0.8$. On the stability curves we have 
used black on green, blue on red, gray on black, red on blue and cyan on 
magenta to highlight the stable WA phases. Note that in the left panel, 
the $\log T$ axis extends up to $9$. The flat spectrum ($\alpha = 0.2$) 
curves have higher temperature Compton stabilization at $T > 10^8 \kel$.}
\labfig{slope}
\end{center}
\end{figure*}


\section{Influence of EUV / Soft X-ray slope}
\labsecn{influencespectral}

In this section we consider the effect on the stability curve of changing the
shape of the continuum in \equn{2power-law} by varying $\alpha$ and $\alpha_{ox}$
over a range of values. All through the section we assume that the gas has
chemical composition of solar abundance.

\subsection{Influence of Soft X-ray slope}
\labsubsecn{slope}

The left panels of \fig{slope} shows the shapes of the ionizing continua 
and the corresponding stability curves for different values of $\alpha$, 
$\alpha_{ox}$ is held at a constant value of $1.2$ for these curves. We have 
highlighted the stable $10^5 \kel < T < 10^7 \kel$ warm absorber states. 

For very flat continua, with energy index $\alpha\sim0.2$ or less, almost no 
stable state exists in the temperature range of $10^5 - 10^7 \kel$, since there 
are no regions of the curve here with positive slope except for a very narrow 
one at $\log(\xi/T) \sim -2.6$. The large number of high energy photons pushes 
the Comptonization temperature to $T > 10^8 \kel$. As noted by Krolik 
$\etal$ (1981), there is a cooler phase at $\log T < 4.5$ and a hot phase which 
is transparent to atomic features. Warm absorbers are thus unlikely to be found 
in AGN which have such flat X-ray spectra. Stability curves for steeper incident 
spectra show additional stable states in the intermediate temperature regime. 
Curves corresponding to $\alpha = 0.5$ and $\alpha=0.8$ are both stable up to 
$\sim2.5 \times 10^5 \kel$. The {\it standard} curve shows one extensive stable 
state at $\sim 10^6 \kel$. We will discuss the effect of multiple phases further 
in \secn{multiphase}. For steeper spectra with $\alpha=1.1$, the absorbing gas is 
predicted to be stable for all $\xi$ with the Compton temperature now lying at 
$\log T =6.9$. The distinct multiple phases at different temperatures, 
but in near pressure equilibrium are lost. For such steep continua, stable states 
exist over a continuous distribution of $\xi/T$ and $T$ in the phase space.  

In summary, AGN having flat X-ray spectra with $\alpha < 0.4$ are not likely to
have WAs. Although a steeper continuum can lead to physical conditions that 
can sustain warm absorbing gas in thermal equilibrium, multiphase nature of WA 
can only be seen for AGN spectra with intermediate slopes of $\alpha \sim 0.8$. 
It is interesting to note that multiphase WAs are predicted for continua 
having slopes in the range $0.7<\alpha<0.9$ which is the observed range for the 
bulk of the observed AGN population (Wilkes \& Elvis 1987; Grupe $\etal$  2006; 
Lopez $\etal$  2006). 

We would like to note an interesting difference between results obtained by us 
and that by Krolik \& Kriss (2001, hereafter KK01). Using a model ionizing 
continuum given by Kaspi \etal (2001), KK01 have obtained stability curves 
computed by version 2.1 of XSTAR (Kallman 2000\footnotemark). The ionizing 
continuum by Kaspi \etal (2001) has values of $\alpha_{ox} \sim 1.3$ and 
$\alpha = 0.77$, very 
similar to those used in the {\it Standard set}. However, KK01 find that the 
corresponding stability curve shows a steep near vertical rise in temperature 
from $\sim 3 \times 10^4 \kel$ to $\sim 10^6 \kel$ for $40 < \xi < 1500$ which 
gives a continuous distribution of stable states over a wide range of $\xi$, 
while pressure ($\xi/T$) remains almost constant. For the same range of $\xi$, 
($50 - 1200$), our {\it Standard} curve, shows two distinct WA phases in 
pressure equilibrium with each other and the change in temperature is gradual 
and spans a much wider range of $\xi/T$ values.
\footnotetext{URL : http://heasarc.gsfc.nasa.gov/docs/software/xstar/docs/html/\\xstarmanual.html.}
%

\subsection{Influence of $\alpha_{ox}$}
\labsubsecn{alphaOX}

In the right panels of \fig{slope} we have shown the results of varying the EUV
to soft X-ray slope $\alpha_{ox}$ (see \equn{alphaOX}). The X-ray slope $\alpha$ 
is held at a
constant value of $0.8$. Observations find $\alpha_{ox}$ between $\sim 1$
and $\sim 2$ as reported by Strateva $\etal$  2005 and Steffen $\etal$  2006. 
They find that there is an anti correlation between $\alpha_{ox}$ and the ultraviolet 
luminosity $L_{\nu}(2500 \AA)$ as was first reported by Tananbaum \etal in 1979. 
We have used the representative 
values of $1.0, 1.4, 1.6 \,\, \rm{and} \,\,1.8$ in addition to the {\it Standard} 
value $1.2$. The WA phases are highlighted as before.

From \fig{slope} we see that the nature of stability in the $\alpha_{ox} = 1$ 
curve is very similar to that in the $\alpha_{ox} = 1.2$ curve. However, beyond 
$1.2$, with a sharper drop in flux from the EUV to the soft X-ray region, \ie 
with steeper value of $\alpha_{ox}$, there is a gradual increase in the stability 
of the absorbing gas at $10^5\kel$, and a gradual shift to higher pressure ($\xi/T$). 
The $\alpha_{OX} = 1.4$ curve shows three 
distinct stable states at $10^5, 10^6 \,\, \rm{and} \,\, 10^7 \kel$ in pressure 
equilibrium with each other. For steeper $\alpha_{ox} = 1.6, 1.8$, stability at 
$10^5 \kel$ is increased. The $\alpha_{ox}=1.8$ curve shows no unstable states. 
For such spectra, the WA can have a continuous distribution of pressure and density. 
The Compton temperature for the stability curves understandably gradually drops 
from $\log T \sim 7.7$ for $\alpha_{OX} = 1.0$ to $\log T \sim 6.0$ for 
$\alpha_{OX} = 1.8$ as the ratio of EUV to X-ray photons increases and Comptonization 
is progressively dominated by lower energy photons.

To summarise, the existence of the WA is independent of the EUV to X-ray 
slope of the ionizing continuum. However, its multiphase nature is influenced 
by this slope; the steeper the fall of flux from EUV to X-ray (\ie larger 
$\alpha_{ox}$, the lesser is the possibility of 
finding a multiphase WA.


\subsection{Influence of $E_{max}$}
\labsubsecn{Emax}

\begin{figure}
\begin{center}
\includegraphics[scale = 1, width = 8.0 cm]{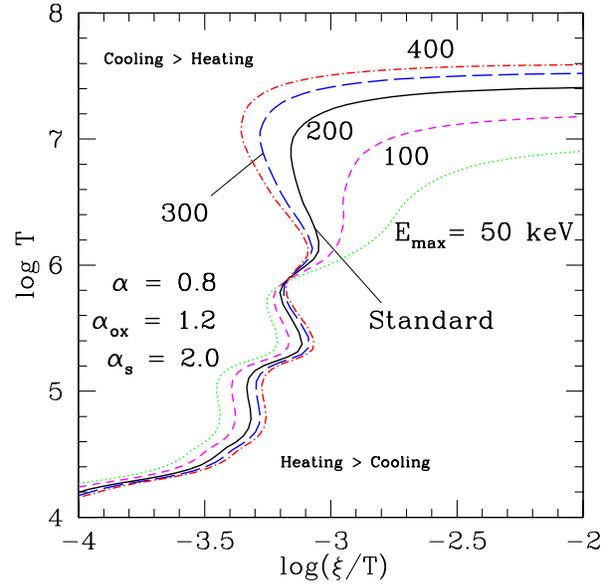}
\caption{Stability curves drawn using different values for the exponential 
higher energy cut-off $E_{max}$. The curves are appropriately labeled. All 
other physical parameters are same as in the {\it Standard set}}  
\labfig{Emax}
\end{center}
\end{figure}

The {\it Standard set} ionizing continuum is exponentially cut off at 
$E_{max} = 200 \kev$ (see \equn{2power-law}). Results from the Swift/BAT 
(Burst Alert Telescope)
hard X-ray sky survey (Tueller et al., 2008) show that, of the
brightest few dozen AGNs, for which $E_{max}$ can be determined, the
lowest value is $40 \kev$, while some 2/3 of the objects have $E_{max} >
120 \kev$, the highest value that BAT can measure. As a result an
appropriate value of $E_{max}$ for the {\it Standard Set} needs to be set
well above the {\it Early Set} value. We adopt $E_{max} = 200 \kev$. The 
high energy 
cut-off is important in Compton processes and hence in determining the Compton 
temperature of a stability curve, but is not likely to affect its WA region. 
However, for the sake of completeness we examine the effect on stability curves 
of varying the value of $E_{max}$.

\fig{Emax} shows the various stability curves drawn using the different values 
for $E_{max}$. The curves are appropriately labeled. All other physical 
parameters are as in the {\it Standard set}. For the lowest considered value 
of $E_{max} = 50 \kev$, the Compton processes start dominating the thermal balance 
at lower temperatures resulting in enhanced stability at $T \sim 10^6 \kel$. 
However, the influence is still not significant enough to change the overall 
qualitative nature of the WA or its multiphase properties. The majority of the 
Swift/BAT AGN have $E_{max} > 100 \kev$. For such values of $E_{max}$, the WA remains 
effectively insensitive to the variation in the high energy cut-off; the only 
influence is in increasing the Compton temperature with the increase in $E_{max}$.


\section{Influence of Density of the absorber}
\labsecn{density}

\begin{figure*}
\begin{center}
\includegraphics[scale = 1, width = 18 cm, trim = 30 180 0 20, clip]{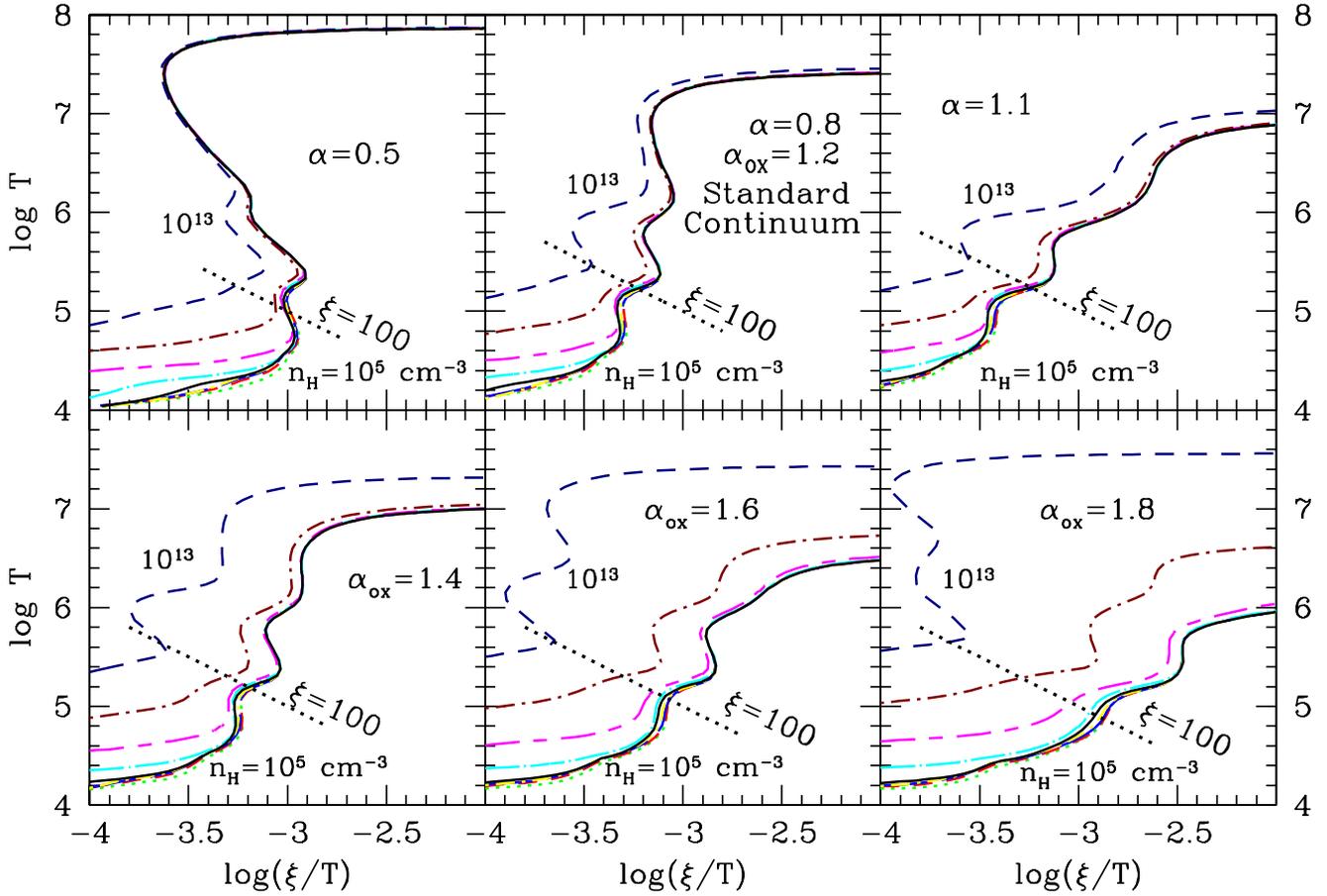}
\caption{The density dependence of the stability 
curves as a function of the ionizing continuum. The top panels correspond to the 
various values of the X-ray slope $\alpha$ whereas the bottom panels are for 
different values of EUV to X-ray slope $\alpha_{OX}$ and all the panels have been 
appropriately labeled. Density at any point of a stability curve is given by 
$n_H = (L/R^2)/\xi$. For each ionizing continuum, several 
stability curves are drawn by varying the ratio 
$L/R^2$ from $10^7 \,\,\rm{to}\,\, 10^15$ in 
multiplicative steps of $10$ such that the density at $\xi=100$ varies from 
$n_H = 10^5 \,\,\rm{to}\,\, 10^{13} \,\,\rm{cm^{-3}}$. The dotted black lines 
cutting across the curves are the constant $\xi = 100$ lines. In the range 
$n_H = 10^5 \,\,\rm{to}\,\, 10^{10} \,\,\rm{cm^{-3}}$, the density value does not 
affect the WA range of the stability curve at all. For $n_H > 10^{10} \,\,\rm{cm^{-3}}$, 
the density dependence of the stability curve is seen to increase with the steepness 
of the ionizing continuum.}
\labfig{density}
\end{center}
\end{figure*}


As stated in \secn{constraints}, we have assumed that the stability curve is 
insensitive to the variation in the value of hydrogen density $n_{\rm{H}}$ of the 
WA gas. Based on this assumption, we have taken the liberty to choose an arbitrary 
fiducial value of $n_{\rm{H}} = 10^9 \,\,\rm{cm^{-3}}$ for $\xi = 100$, the typical 
value of ionization parameter reported for WAs, thus yielding 
$L/R^2 = \onlyten{11}$ from \equn{ionparm}, which we hold constant throughout this 
paper except in this section. Many authors carrying out 
photoionization calculations make similar assumptions (Hess, Kahn \& Paerels, 1997; 
Krolik, McKee \& Tarter, 1981; Krolik \& Kriss, 2001, RF95) and RF95 have even 
verified this statement over a wide range of densities 
$10 \le n_{\rm{H}} \le 10^{11} \,\,\rm{cm^{-3}}$ for a simple power-law ionizing 
continuum with $\alpha = 0.8$. However, more recently, 
Rozanska, Kowalska \& Goncalves (2008) showed that 
the stability curve shows remarkable density dependence if the ionizing continuum 
has a complicated shape with relatively more soft X-ray, photons although the same 
exercise with a simple power-law ionizing continuum with $\alpha = 1.0$ yields no 
density dependence on the part of the stability curve. In this paper we have studied 
the stability curve's behaviour as a function of complicated ionizing continua with 
having different ratios of the soft to hard X-ray photons. Hence it is an interesting 
exercise to check if the stability curves, especially in the range 
$10^5 < T < 10^6 \kel$, drawn using the different ionizing continua considered in 
this paper show any variation if the density is changed.

In this section we relax the assumption made in \secn{constraints} that 
$L/R^2 = \onlyten{11}$ and for each ionizing continuum considered in 
\secn{influencespectral}, draw several stability curves by varying this ratio from 
$10^7 \,\,\rm{to}\,\, 10^{15}$ such that the density at $\xi=100$ varies from 
$n_{\rm{H}} = 10^5 \,\,\rm{to}\,\, 10^{13} \,\,\rm{cm^{-3}}$ in multiplicative steps 
of ten from one curve to another. The results are shown in \fig{density}. The upper 
panels 
in the figure show the stability curves for continua with different values of the 
X-ray slope $\alpha = 0.5, 0.8 \,\,\rm{and}\,\,1.1$, whereas the bottom panels 
correspond to different values of the EUV to X-ray slope 
$\alpha_{OX} = 1.4, 1.6 \,\,\rm{and}\,\, 1.8$. The determination of thermal equilibrium 
of the WA is seen to be independent of density in the range 
$10^5 \le n_{\rm{H}} \le 10^{10} \rm{cm^{-3}}$, whatever be the shape of the ionizing 
continuum. Stability curves with  
$n_{\rm{H}} = 10^{11} \,\,\rm{and}\,\, 10^{12} \,\,\rm{cm^{-3}}$ show insignificantly 
small differences from the rest of the curves with lower densities as we vary $\alpha$ 
from 0.5 through 0.8 to 1.1 This indicates that the soft X-ray slope has little or no 
role to play in inducing density dependence to the estimations of thermal equilibrium 
conditions. The most interesting result of this exercise is that as the EUV to X-ray 
slope $\alpha_{OX}$ is increased from 1.2 through 1.4 and 1.6 to 1.8, the WA thermal 
equilibrium properties become significantly density dependent for 
$n_{\rm{H}} \ge 10^{11} \,\,\rm{cm^{-3}}$ and the dependence is more pronounced with 
the increase in $\alpha_{OX}$ i.e. with the increase in the relative number of EUV 
photons. $n_{\rm{H}} \sim 10^{12} \,\,\rm{cm^{-3}}$ can be considered to be an upper 
limit on WA density and would be consistent with clouds residing within the $\hbeta$ 
broad emission line region, on the scale of the accretion disc, a few thousand 
Schwartzchild radii (Krongold $\etal$, 2007). Thus higher values of density are 
unphysical for WA. However, in each of the panels in \fig{density}, we have also 
plotted the $n_{\rm{H}} \sim 10^{13} \,\,\rm{cm^{-3}}$ curve for comparison and it 
shows gross difference from the stability curves with lower densities for all the 
continuum shapes considered.

Since the $\alpha_{OX} = 1.8$ curves show the maximum sensitivity to density 
variation, we select them to investigate the cause of this effect. We choose the 
$\xi = 100$ model from each of the curves with densities 
$10^5 \le n_{\rm{H}} \le 10^{13} \,\,\rm{cm^{-3}}$ and check what are the dominant 
heating and cooling agents in determining the thermal balance. The fractional 
contribution towards the total cooling rate by the dominant cooling agents undergoes 
only a gradual change as we change the density though the above mentioned range. 
The remarkable change is shown by the dominant heating agent. For 
$n_H \le 10^{10}\,\,\rm{cm^{-3}}$, the heating agents are all photoionized species 
and free-free heating starts to dominate for $n_H = 10^{11}\,\,\rm{cm^{-3}}$. The 
percentage of heating by free-free absorption increases from $\sim 17$ for 
$n_H = 10^{11}\,\,\rm{cm^{-3}}$ to $\sim 83$ for $n_H = 10^{13}\,\,\rm{cm^{-3}}$.

The ionization equilibrium conditions of the absorbing gas can be uniquely 
determined if the shape of the ionizing continuum and the flux received by the gas 
is specified. Assuming that the absorber is in ionization equilibrium, {\tiny CLOUDY} 
evaluates the heating and cooling rates at different temperatures for the given gas 
and determines the thermal equilibrium solution for the temperature where these two 
rates balance each other. In the WA temperature regime where photoionization is the 
dominant heating process and thermal equilibrium is achieved by recombination cooling, 
the temperature estimates become independent of density as both the heating and the  
cooling terms have similar dependence on density which cancels out from both sides of 
the thermal balance equation. At higher temperatures, the thermal balance is attained 
by Compton heating and cooling, in which case also, the density cancels out. 

However, the interplay between various heating and cooling 
processes is also a function of the ionization continuum. For spectral shapes with 
steep $\alpha_{OX}$ there maybe sufficiently large number of low energy photons to 
make the free-free absorption by ionized hydrogen dominate over photoionization as the 
heating process. The heating rate for this interaction has a different dependence on 
density and will result in making the thermal equilibrium conditions 
sensitive to density variations for $n_{\rm{H}}$ greater than a certain threshold value.

The interesting point to note at the end of this exercise is that although the mode of 
photoionization calculations used by us and Rozanska \etal (2008) are different, our results 
are qualitatively consistent with each other. For sufficiently soft UV dominated ionizing 
continuum the thermal equilibrium calculations become density dependent if the gas is 
denser than a certain threshold. Since the gas is also assumed to be in ionization 
equilibrium, the ionization structure will also be influenced. The effects would be much 
more pronounced for {\it constant pressure} mode of calculations as shown by 
Rozanska \etal (2008). 
The importance of this result is that for dense gas illuminated by UV dominated 
ionizing continua, the density of the gas can be estimated directly by modeling the 
observed spectra in UV.


\section{Influence of the Chemical Composition of the Absorber}
\labsecn{chemicalcomp}

\begin{figure}
\begin{center}
\includegraphics[scale = 1, width = 8 cm, trim = 5 0 175 0, clip]{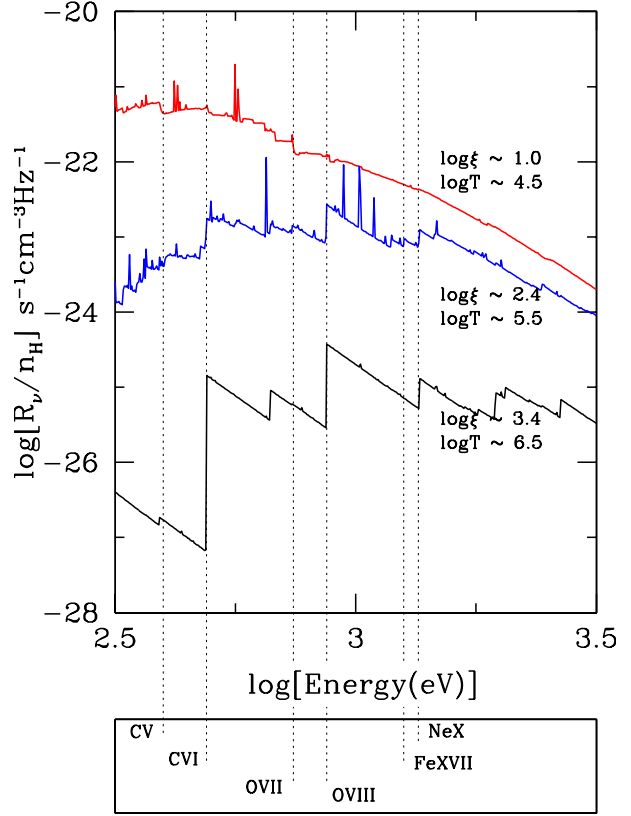}
\caption{The radiation field interaction rate is plotted as a function of energy
for physical models with different ionization parameters 
chosen from different regions of
the standard stability curve. See text for details. The dotted vertical lines
mark the absorption edges of some of the ions important for WA,
viz. CV, CVI, OVII, OVIII, FeXVII, NeX.}  
\labfig{iongas}
\end{center}
\end{figure}


The behavior of the models along the stability curve, from $\log T \sim 4.0$ to
$\log T \sim 7.0$ is sensitive to the role of atomic physics in determining the
thermal state of the gas.  The photoionization cross section 
$a_{\nu}(X^{+i})$ from the ground level of $X^{+i}$ ion with the threshold 
$\nu_i$ is given by the equation
\begin{eqnarray}
n(X^{+i})\,\Gamma(X^{+i}) & = & n(X^{+i})\,\int ^{\infty}_{\nu_i}\,\frac{4\,\pi\,J_{\nu}}{h\,\nu}\,a_{\nu}(X^{+i})\,d\nu \nonumber \\
& = & n(X^{+i})\,\int ^{\infty}_{\nu_i}\,R_{\nu}(X^{+i})\,d\nu. 
\labequn{iongas}
\end{eqnarray}
where $n(X^{+i})$ is the number density of the $X^{+i}$ ion, $\Gamma(X^{+i})$ is
the number of excitations per particle to the $(i+1)^{th}$ state and $J_{\nu}$ is
the mean intensity. Thus the function $R_{\nu}(X^{+i})$ is a measure of the 
interaction between the radiation field and the $X^{+i}$ ion.

\begin{figure*}
\begin{center}
\includegraphics[scale = 1, width = 19 cm, trim = 10 425 5 125, clip]{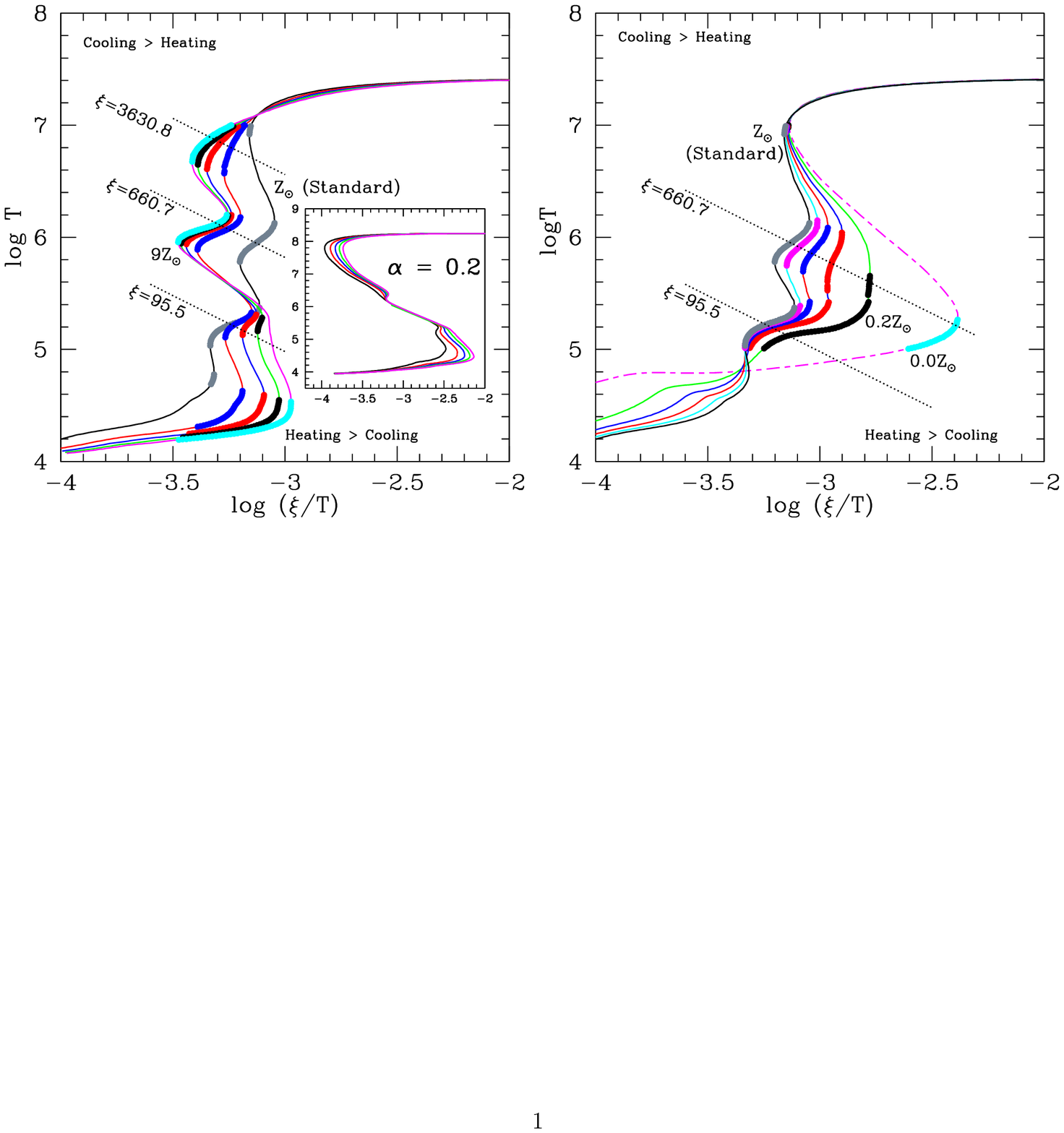}
\caption{Stability curves for absorbers with different abundances. The left 
panel shows curves having abundance from $\zsol$ to $9\zsol$ in steps of 2.  
The curves show that absorbers with super-Solar chemical compositions will 
enhance multiphase nature of the absorber. Except for the curves in 
the inset window in the left panel, the {\it Standard set} ionizing 
continuum with $\alpha = 0.8$ has been used for the calculations. The 
stability curves in the inset window are, however, obtained using an 
ionizing continuum with flat X-ray slope of $\alpha = 0.2$ to investigate 
if enhanced metallicity helps sustain WA for gas irradiated with such flat 
continuum. The results show no significant stable states in the temperature 
range $5 < \log T < 7$ for WAs. The right panel shows the curves having 
sub-Solar abundances from $0.0\zsol$ to $\zsol$ in steps of 0.2. 
Decreases in metallicity are seen to stabilize 
the curve in the temperature range of $10^5 \,-\,10^6 \,\kel$. However, the 
possibility of multiphase media decreases with the 
decrease in the abundance. In all the panels the stable WA are highlighted 
as before. The dotted black lines cutting across the stable WA states are 
constant $\xi$ lines. $\xi = 95.5$ and $\xi = 660.7$ correspond to the 
middle of the WA states of the {\it Standard} curve at 
$10^5 \kel$ and $10^6 \kel$ respectively and $\xi = 3630.8$ corresponds to 
the middle of the $10^7 \kel$ phase for the $Z/\zsol = 3$ stability curve.}
\labfig{abundance}
\end{center}
\end{figure*}

In \fig{iongas} we have shown the dependence of 
$$R_{\nu} = \sum \limits_i R_{\nu}(X^{+i})$$ 
on $\xi$ and hence 
on the ionization state of the absorbing gas. We have selected three different 
models along the {\it Standard} stability curve and plotted 
$R_{\nu}/n_{\rm{H}}$ 
as a function of energy with the values of $\log\xi$ and $\log T$ indicated for 
each curve; $n_{\rm{H}} = 10^{11}/\xi$ being the hydrogen density. The
absorption edges of various important ions including CV, CVI, OVII, OVIII,
FeXVII and NeX which are the signatures of WA have been indicated with dotted
lines. As the gas becomes more ionized  by factors of $\sim 10$ its interaction 
with radiation decreases by orders of magnitude, especially at energies less than 
$1 \kev$. At any given energy the value of $R_{\nu}$ is lower for higher $\xi$. 
However, with highly ionized gas, the interaction shifts to the higher energies. 
For $\log \xi = 1$, there is no sharp increase in $R_{\nu}$ at energies 
corresponding to the WA edges; these interactions grow with $\xi$ 
as seen in the middle and the bottom curves corresponding to $\log \xi = 2.4$ 
and $3.4$ respectively.

We see above that at values of $\log \xi$ relevant for WAs, the nature of the 
interaction between gas and radiation is significantly influenced by the atomic 
physics of the heavier elements. In the following subsections we have studied the 
effect of different chemical compositions of the absorber 
on the stability curve. The illuminating continuum is assumed to be the same 
as that in the {\it Standard set} values (Table 1). We have generated the 
stability curves for (i) super-Solar abundances, (ii) sub-Solar abundances and 
(iii) other abundances where specific elements or group of elements have been 
removed from the solar metallicity.

\begin{figure*}
\begin{center}
\includegraphics[scale = 1, width = 19 cm, trim = 10 425 5 125, clip]{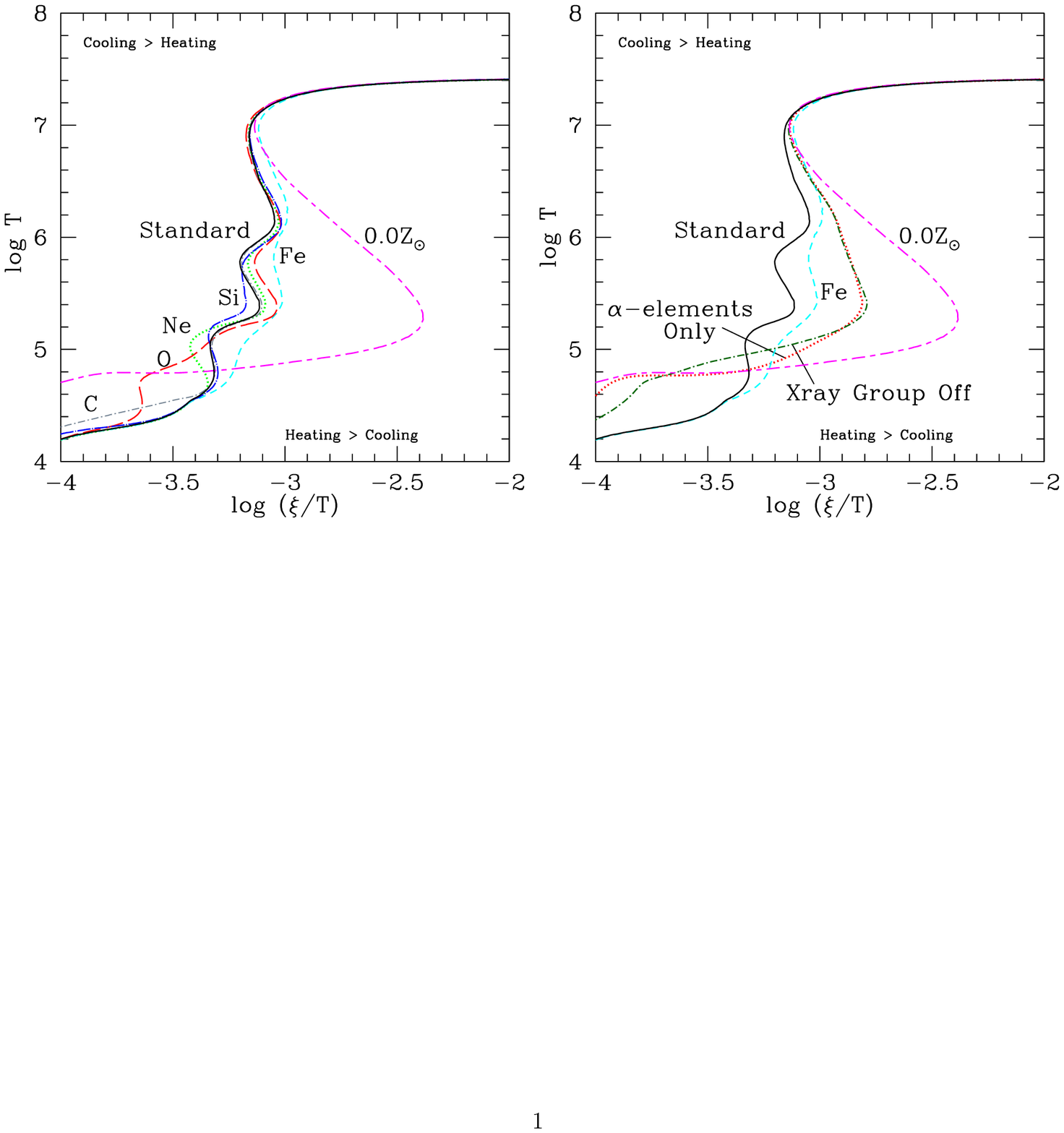}
\caption{Stability curves for different chemical
compositions with certain individual elements or certain groups of elements, as labeled, 
set to zero. "X-ray Group Off" signifies the absence of the group of metals C,O, Ne, and Fe 
in the gas abundance, whereas "$\alpha$-elements Only" represents an absorbing gas having only 
$\alpha$-elements (Ne, Mg, Si, S, Ar, Ca, Ti) as heavy elements. See text for further details. 
The curve labeled `Standard' in the two panels uses the {\it Standard set} of 
values (\tablem{table2}). We have also plotted the {\it zero metals} curve in 
both the panels for comparison}
\labfig{elementsoff}
\end{center}
\end{figure*}

\subsection{Super-Solar Abundance}
\labsubsecn{supersolar}

The left panel of \fig{abundance} shows stability curves for gas having super-Solar 
abundances as defined in Hamann \& Ferland (1993). In addition to the $10^4\kel$ and 
the $10^8 \kel$ phases, all the curves also exhibit WA states. However the 
temperatures of the WA phases and their distribution in the phase space change with 
metallicity. The nature of stability exhibited can be separated into three different 
groups showing distinctly different trends.

All the curves in the group with solar 
to three times solar exhibit stable gas at 
$T\sim10^5\kel$, with the extent of the stable region decreasing as the metallicity 
is increased. $\xi_5 = 95.5$ for the {\it Standard} curve, $\xi_6$ being the ionization 
parameter corresponding to the middle of the $10^5 \kel$ WA phase. The range of ionization 
parameters for the $10^5 \kel$ gas is $50 \le \xi \le 182$ for the {\it Standard} curve. 
The $\xi = 95.5$ drawn in the left panel of \fig{abundance} indicates that the $10^5 \kel$ 
WA is pushed towards higher values of $\xi$ as the metallicity increases. 
Another stable states appears in 
all the curves at 
$T\sim\,\,10^6\,\kel$, but this time with the increase in metallicity, the extent of 
the stable region is increased. $\xi_6 = 660.7$ for the {\it Standard} 
curve, where $\xi_6$ is defined to be the ionization parameter corresponding to the 
middle of the $10^6 \kel$ stable WA phase which shows the range 
$380 \le \xi \le 1202$ for the {\it Standard} curve. The solid black line across the 
$10^6 \kel$ phases of the different stability curves is the $\xi = 660.7$ line. 
Thus, we see that with the increase of 
metallicity, $\xi_6$ is decreased slightly and the $10^6 \kel$ stable phase moves to 
lower $\xi/T$ and higher $T$. A very narrow (in the range of $\log(\xi/T)$) 
phase is seen to appear at $T\sim\,\,10^7\,\kel$ for metallicity of 3 $\times$ 
solar. The detailed quantitative comparisons are made in \secn{multiphase}. 

The curves in the next group with $4 \le Z/\zsol \le 7$, are very similar to each 
other, with a very narrow stable state at $10^5\kel$, which is not in pressure 
equilibrium with the $10^6\kel$ phase and the $10^7\kel$ state becomes more prominent. 
As abundances are increased further ($Z/\zsol > 7$) the stable state at 
$10^5\kel$ disappears, while the $10^6\,\kel\,\,\rm{and}\,\,10^7\kel$ states remain 
quite wide and in pressure equilibrium with one another. The stability curves for 
the super-Solar abundances thus suggest that the possibility for the warm absorbing 
medium to exist in discrete phases in pressure equilibrium with each other increases 
with increasing metallicity. We return to these issues in \secn{multiphase} 
with quantitative details.

The stability curves for the super-Solar metallicities with $Z/\zsol > 3$ are 
seen to have gas at $T \le 6 \times 10^4 \kel$ to be in pressure equilibrium with the 
WA states at higher temperatures. Such stable states at $\sim 10^4 \kel$ are also 
highlighted in the corresponding stability curves in the left panel of \fig{abundance}. 
Gas at $\sim 10^4 \kel$ might be representing the `BELR clouds'. Hence this phenomenon 
might indicate the interesting possibility of some connection between the X-ray and UV 
absorber, as discussed by Turner $\etal$  (1995) and Elvis (2000). However, these 
results have to be checked out further with more realistic AGN spectra which is beyond 
the scope of this paper and will be addressed in our next publication 
(Chakravorty \etal, in preparation).

We have also investigated the possibility of high metallicity gas making it possible 
for WA to exist even when the X-ray ionizing continuum is very flat. 
The stability curves are shown in the small inset window of the left panel of 
\fig{abundance}. Even for solar abundance, the  $\alpha=0.2$ curves did not 
show any significant $10^5 \kel$ WA and higher metallicity diminishes the chances 
further. At higher temperatures of $\sim 10^6 \kel$, very narrow 
regions of stability appear for $Z/\zsol \ge 5$. However, even in these cases there 
is no multiphase exhibited by the stability curves. Thus even if the gas has high 
metallicity, WAs are not likely to be present in the intervening medium, if the 
ionizing continuum is very flat.


\subsection{Sub-solar abundance}
\labsubsecn{subsolar} 

Gas in AGN has super-Solar metallicity (Hamann \& Ferland, 1999). However, AGN 
illuminate low density gas
on large scales which has low abundances, especially at high redshift. Sub-Solar
chemical composition are relevant for such line of sight absorbing systems
including Damped Lyman-$\alpha$ systems in the inter galactic medium and smaller
galaxies like the Magellanic clouds. The inter-cluster medium, for example, has
$\sim0.3\zsol$ abundance and in certain systems there are even stars with an
abundance of $0.01\zsol$ (Pettini, 2006 and references therein). 
It is, therefore, interesting to examine the effects
of sub-Solar abundance on the stability curve. We show the low metallicity 
stability curves in the right panel of \fig{abundance}. As before, the 
solid black lines across the stable WA phases are the constant $\xi$ lines.

The ionization parameter $\xi_5$ corresponding to the middle of the $10^5 \kel$ 
stable phase increases from its standard value of $\xi_5 = 95.5$ with the 
decrease in metallicity and there is gradual enhancement of the thermal stability 
of gas. On the other hand, with almost constant $\xi_6 = 660.7$, the stable 
phase at $\log T \sim 6$ moves towards lower temperature and higher pressure 
($\xi/T$) and its extent in the 
$\log (\xi/T)$ shrinks until such a phase disappears for $0.0 \zsol$. Detailed
quantitative comparisons will be presented in \secn{multiphase}.

Absorbing gas with {\it zero metals} might be relevant for super massive black 
holes (SMBHs) or their seeds at redshift $z > 6$, which are formed by direct 
collapse and where no star formation precedes the radiating SMBH 
(Volonteri \& Rees, 2005; Begelman, Volonteri \& Rees, 2006). We have shown the 
{\it zero metals} stability curve in the right panel of \fig{abundance}. There 
is a single phase WA at $10^5 \kel$ with significantly high $\xi_5 =437$ as 
compared to the {\it Standard} $\xi_5 = 95$.

\subsection{Influence of individual elements and groups of elements}
\labsubsecn{elemetsoff}

\fig{iongas} has shown how the interaction between gas and radiation is 
sensitive to the ionization potentials of various ions of elements like C, O, 
Fe and Ne indicating that the WA states are likely to be affected not only by 
the overall abundance of the absorber, but also by specific elements and groups 
of elements. \fig{elementsoff} shows the effect on the stability curves if certain 
elements or groups of elements are not present in the absorbing gas. 
In both the panels of the figure, we have also drawn the stability 
curve for the {\it Standard set} of parameters and {\it zero metals} 
for comparison.

The WA consists of absorption edges and lines of different 
ionization stages of oxygen, carbon, neon and iron. We define the 
{\it X-ray group} constituting of only these four elements. The left
frame of \fig{elementsoff} shows the results of simulating various stability
curves corresponding to the removal of different individual elements of the
group from a solar composition gas. Setting carbon abundance to zero 
does affect the stability curve at $\log T < 4.5$ and the deviations imply 
that Carbon acts as a cooling agent at those temperatures. However at higher 
temperatures relevant for WA, Carbon has no significant effect. The removal 
of oxygen and neon
leads to deviations in the stability curves which indicate that for 
$T < 10^5 \kel$ oxygen acts as a significant cooling agent with some 
contributions from neon, whereas for $T > 10^5 \kel$ both act as heating 
agents. Removal of iron has the maximum effect as it 
renders the curve relatively featureless in 
the range $4.5 \le \log T \lesssim 7$. Such an effect is expected because at 
higher energies iron is the most abundant absorbing element and plays a major 
role to stabilise the gas by acting as a significant heating agent. This result 
has interesting cosmological consequences. Type Ia supernovae created iron when 
the Universe was about 1 Gyr old (Hamann \& Ferland, 1999 and the references 
therein). Hence there would be an interesting transition in the nature of the WA 
at this age; WAs devoid of iron from before this era cannot show a significant 
$10^6 \kel$ phase. Future missions like Constellation X might be able to detect 
such a change. Like the removal of iron, the removal of the {\it X-ray group} 
(O, C, Ne, Fe) as a whole also {\it flattens} out the stability curve, but to a 
larger extent as shown in the right panel of \fig{elementsoff}. If the 
absorbing system is devoid of the {\it X-ray group}, the stable states at 
$10^5 \kel$ are pushed towards higher values of $\xi$ with $\xi_5 = 158$ as 
compared to $\xi_5 = 95$ for the {\it Standard} curve and there are no higher 
temperature stable states thus nullifying the possibility of multiphase WA.

In addition to the elements of the {\it X-ray group}, it is instructive to 
study the effect of the $\alpha$-elements (Ne, Mg, Si, S, Ar, Ca, Ti) 
produced together with oxygen in type II supernovae which precede the type 
Ia supernovae where iron and associated elements are formed. Thus the 
relative abundance of these elements gives a handle on the stellar 
evolution history and the age of the galaxy. Ne is the only $\alpha$-element 
which has been studied as a part 
of the {\it X-ray group} (O, C, Ne, Fe) as well, and its effect on the stability 
curve has already been discussed above. Among the remaining $\alpha$-elements, 
magnesium, sulphur, argon, calcium and titanium have negligible effects. 
Only silicon is a 
significant cooling agent at $\log T \sim 5$, so that removal of silicon shows 
an almost vertical jump from $10^5 \kel$ to $10^6 \kel$ in 
the corresponding stability curve in the left panel of \fig{elementsoff}. 
As a group however, $\alpha$-elements suggest interesting consequences. 
According to the evolution history of metals in the Universe, 
the $\alpha$-elements are formed first, followed by the other metals in 
subsequent stages. Hence we 
investigated the nature of an absorbing gas which has only the $\alpha$-elements 
as metals and the corresponding stability curve is shown in the right panel of 
\fig{elementsoff} labeled as `$\alpha$-elements Only'. For $\log T \ge 5.2$ the 
curve is same as the ``{\it X-ray group} removed'' curve and at lower temperatures, 
$\log T \sim 4.5$, it traces the $0.0 \zsol$ curve. These results suggest that 
a certain amount of evolution of metals is necessary for any absorbing systems 
to show WA signatures and might set an upper limit to the redshifts at which 
warm absorbers can form.  

\begin{figure}
\begin{center}
\includegraphics[scale = 1, width = 8 cm]{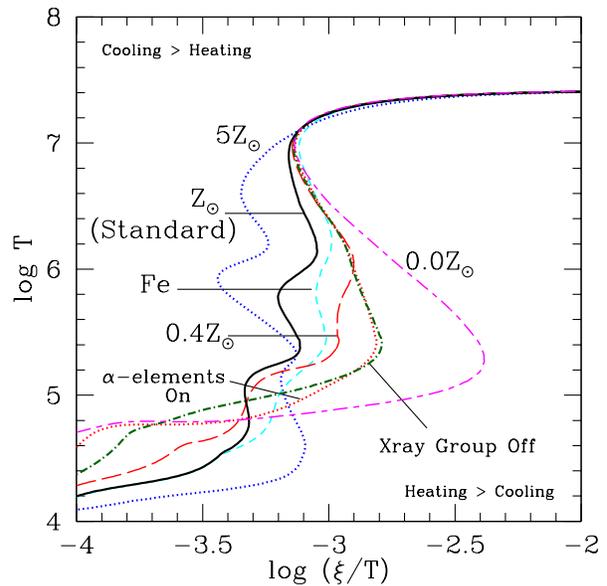}
\caption{A summary plot depicting the stability curves for a range of chemical
compositions. The curves marked `Fe' and `X-ray Group Off' correspond to absorbers
where iron and the X-ray group elements respectively are absent. An absorber with 
a composition where only the $\alpha$-elements are present as the metals, is represented by 
the stability curve labeled with `$\alpha$-elements On'. There is a
gradual flattening of the stability curve as one moves from $5\zsol$ to
$0.0\zsol$.}  
\labfig{abundancesummary}
\end{center}
\end{figure}


\subsection{Summary of abundance effects on stability curves}
\labsubsecn{ElemetsSummary}

In \fig{abundancesummary} we have collected the curves showing the 
major effects due to varying the chemical composition of the absorber. 
As the metallicity of the absorber is decreased
from $5\zsol$ to $\zsol$ and to $0.4\zsol$, the stability curve flattens
gradually. The possibility of finding multiphase WA decreases with
decrease in metallicity. This result is in conjunction with those found by 
Komossa and Mathur (2001). It also decreases if the absorbing medium is deficient
in some elements. If a WA lacks iron, but has all
the other elements with solar abundance, it then has nearly the same stability
properties as an absorber with half solar abundance. The X-ray group of
elements are the most influential in making multiple phases possible. 
Absorbing gas at early epochs of star formation when only 
$\alpha$-elements formed show the same thermal properties as absorbers 
devoid of the X-ray group of
elements, thus suggesting that the properties of WA are significantly dependent 
on the evolution history of metals in the Universe. 
\fig{abundancesummary} also shows the stability curve
for $0.0\zsol$; such a primordial gas of hydrogen and helium has a single 
phase of absorbing gas at $10^5 \kel$ for a significantly high ionization 
parameter. No stable state at $10^6 \kel$ and hence no multiphase solutions 
are possible if the metallicity of the absorber is $< 0.2 \zsol$. Metals thus 
seem to be necessary for absorbing gas to exist at $10^5 \kel < T < 10^7 \kel$. 

Hess \etal (1997) have carried out an extensive study trying to account for the 
unstable thermal equilibrium zones in temperature ranges relevant for AGN and 
X-ray binaries paying attention to even individual ions of various metals 
responsible for causing the instabilities. They have predicted the ionization 
structure and thermal states of absorbers with a wide variety of chemical 
compositions. The results posed in this paper especially in \subsecn{elemetsoff} 
agree qualitatively with their results but as expected, they vary in quantitative 
details. For example, like them, we also find oxygen and iron to be playing 
crucial roles in the ionization and thermal balance equations. However, unlike 
Hess \etal in this paper we find that oxygen and iron influence distinctly 
different temperature zones of the stability curve. As mentioned in their paper, 
such minute details would crucially depend upon the accuracy of the atomic 
physics incorporated which is definitely addressed much better in the current 
photoionization codes than in 1997.

The determination of the particular ions responsible for the heating 
or cooling of the gas at WA temperatures, is very 
sensitive to the exact shape of the ionizing continuum between 50 to 500 eV. Such 
investigations and quantitative comparisons are beyond the scope of this paper 
and will be addressed in our subsequent publications (Chakravorty \etal, in 
preparation).

\section{Multiphase Analysis}
\labsecn{multiphase}

The equilibrium curves we have derived in the previous sections 
often contain segments that allow phases at different temperature to occur 
at similar pressure $\xi/T$. This has clear physical interest as 
distinct phases in pressure equilibrium have been derived in a number of 
cases of WAs 
(Krongold $\etal$, 2003, 2007, Netzer $\etal$, 2003), while in others it 
has been claimed that there is a continuous range of ionization 
parameter (Ogle $\etal$, 2004, Steenbrugge $\etal$, 2005). 
In this section we quantify the extent of these multiphase regions. However, 
we would like to note that for carrying out such an exercise, the atomic 
database needs to be accurate. As mentioned a number of times in the previous 
sections, care has been taken to upgrade the underlying atomic physics in 
{\tiny CLOUDY} with the most recent updates.

\begin{figure*}
\begin{center}
\includegraphics[scale = 1, width = 16 cm, trim = 70 50 70 100, clip]{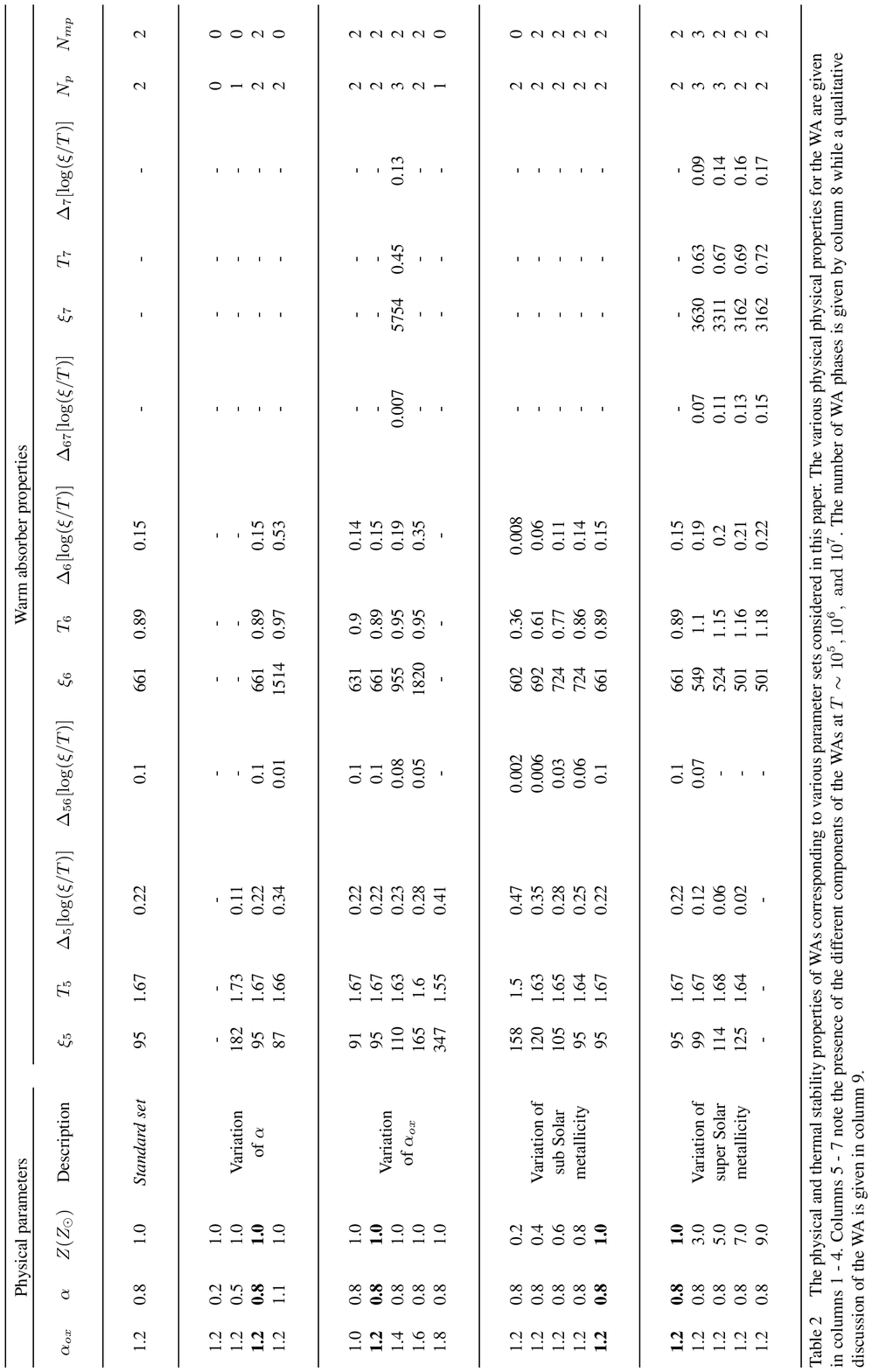}
\end{center}
\end{figure*}

In \tablem{table2} we present detailed quantitative estimates of 
the WA properties for the various parameter sets used in this paper. For 
convenience of representation we introduce a few definitions for use 
in \tablem{table2}. The first four columns describe the 
parameter set used to derive the stability curve and the WA properties. 
$\xi_n$ in columns 5, 9 and 13 and $T_n$ in columns 6, 10 and 14 respectively 
are the ionization parameter and the temperature corresponding to the middle 
of the $10^n \kel$ stable WA phase, where $T_n$ is expressed in units of 
$10^n \kel$. $\Delta_n[\log(\xi/T)]$ in columns 7, 11 
and 15 give the range of $\log(\xi/T)$ over which stable states of WA are 
exhibited for a temperature of $10^n \kel$. For WA to exist in discrete 
multiple phases, different temperature stable states require to exhibit the 
same values of $\xi/T$. The larger the range of these common values, higher 
we presume to be the probability of a multiphase WA. Such intersection in the 
range of $\xi/T$ is represented by 
$\Delta_{56}[\log(\xi/T)]$ for $10^5 \kel$ and $10^6 \kel$ phases and by 
$\Delta_{67}[\log(\xi/T)]$ for the states at $10^5 \kel$ and $10^6 \kel$ and 
their values are noted in columns 8 and 12 respectively. $N_p$ in the second 
last column is the number of stable states in the stability curve in the 
temperature range $5 \le \log T \le 7$ and the number of states in pressure 
equilibrium with each other is given by $N_{mp}$ in the last column. We would 
note that the values of the various quantities recorded in 
\tablem{table2} are useful for understanding trends in the 
variation of the physical quantities as the parameter sets are changed. 
However, for more robust estimations of these numbers and specific 
predictions of the WA properties, we require more detailed modeling of the 
ionizing continuum and the gas abundances. Such investigations will be 
described in a subsequent publication (Chakravorty \etal in preparation).

In \fig{multiphase}, we plot $\Delta_n[\log T]$ against 
$\Delta_n[\log (\xi/T)]$ for a range of continuum shapes (left) and abundances 
(right), where $\Delta_n[\log T]$ and $\Delta_n[\log (\xi/T)]$ are the range 
of $T$ and $\xi/T$ for the stable states at $\sim 10^n \kel$. Presumably, the 
higher the 
values of these two quantities, the better are the chances of existence of 
the state at $10^n \kel$, because any perturbed gas is more likely to find 
a larger stable region.
The various parameter sets used to describe WA conditions in this 
paper have been represented by different symbols which have been 
explained in the bottom panel. The gray square represents 
the {\it Standard set} of parameters and the rest of the
symbols indicate the changes from it. 

\begin{figure*}
\begin{center}
\includegraphics[scale = 1, width = 15 cm]{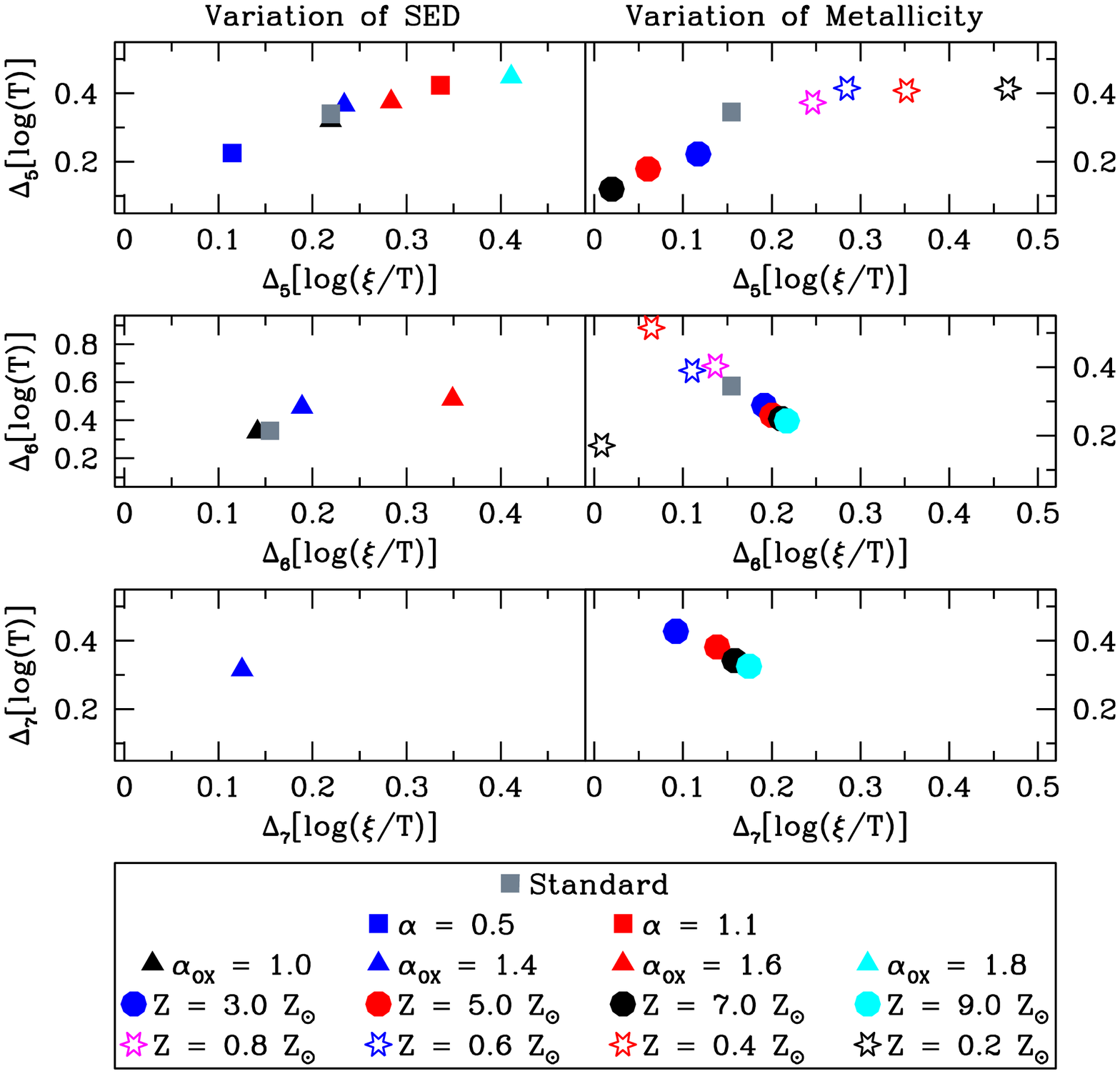}
\caption{ For the stable states of the WAs we have plotted 
$\Delta[\log T]$ against $\Delta[\log (\xi/T)]$ and have 
considered the parameter sets where we have varied the SED of 
the ionizing continuum and the metallicity of the absorbing gas. 
See text for details. 
Various symbols used 
to identify the different parameter sets used in this paper, 
are explained in the bottom panel. The gray 
square represents the {\it Standard parameter set} and for 
the others the deviations from that set are mentioned.}
\labfig{multiphase}
\end{center}
\end{figure*}

From the systematic analysis we can see the following trends:
\begin{enumerate}
\item[$\bullet$] The variation of soft X-ray slope $\alpha$ of the ionizing continuum shows that WAs exist only if $\alpha > 0.2$ and will have higher temperature gas if $\alpha > 0.5$.
\item[$\bullet$] Increases in the soft X-ray slope $\alpha$ decrease the ionization parameter of the $10^5 \kel$ WA while decreasing its temperature, but increases both $\xi$ and $T$ for the $10^6 \kel$ phase. Both $\Delta_5[\log(\xi/T)]$ and $\Delta_6[\log(\xi/T)]$ increases with the value of $\alpha$, which is also demonstrated by the {\it square} data points in the top and the middle left panels of \fig{multiphase} accompanied with increase in $\Delta_n[\log T)]$, indicating that WA phases stabilise with steeper soft X-ray continua.    
\item[$\bullet$] The Multiphase scenario, however, is not facilitated significantly with the variation in $\alpha$. Multiphase solutions at $10^5 \,\, \rm{and} \,\, 10^6 \kel$ are seen only for the standard value of $\alpha = 0.8$. Although the $\alpha = 1.1$ stability curve also shows 2 phases, the value of $\Delta_{56}[\log(\xi/T)]$ is very low, making the possibility of multiple phases in equilibrium seem unlikely.   
\item[$\bullet$] We have found very interesting difference between our results and those obtained by Krolik \& Kriss (2001). They have drawn stability curves for ionizing continuum having properties similar to our {\it Standard set} SED having $\alpha_{ox} \sim 1.2$ and $\alpha = 0.8$. The stability curves shown in KK01 have a near vertical jump in temperature from $\sim 3 \times 10^4$ to $\sim 10^6 \kel$ as $\xi$ runs from $\simeq 40$ to $\simeq 1500$ thus exhibiting a continuous distribution of stable states in $\xi/T - T$ space. For the same set of parameters (ie, the {\it Standard set}) we not only get two distinct WA phases in pressure equilibrium with each other, but also do not see any drastic change in temperature over a narrow range of $\log(\xi/T)$, although our $\xi$ values also run from $\simeq50$ to $\simeq 1200$.
\item[$\bullet$] As the value of the UV-X-ray slope $\alpha_{ox}$ is increased, the ionization parameter for the $10^5 \kel$ WA increases steadily accompanied by a small decrease in temperature. $\xi_6$ also increases but with a small increase in temperature.
\item[$\bullet$] The range $\Delta_5[\log(\xi/T)]$ remains almost constant for $1 \le \alpha_{ox} \le 1.6$, but abruptly increases for the value of 1.8. $\Delta_6[\log(\xi/T)]$ grows abruptly for $\alpha_{ox} = 1.6$ before which it increases steadily but slowly with the increase in $\alpha_{ox}$. There is no significant WA state at $10^6 \kel$ for $\alpha_{ox} = 1.8$, Comptonization becomes the dominant heating and cooling processes at these temperatures for such an SED. The triangles in the left panels of \fig{multiphase} show these trends.   
\item[$\bullet$] Multiphase solutions at $10^5 \,\,\rm{and}\,\, 10^6\kel$ are seen in the range $1 \le \alpha_{ox} \le 1.6$. For $\alpha_{ox} = 1.4$, there is an additional stable state at $10^7 \kel$, but $\Delta_{67}[\log(\xi/T)]$ is very low to claim any possibility of this state being in pressure equilibrium with the lower temperature phases.
\item[$\bullet$] Through the increase of metallicity of the absorbing gas from $0.2 \zsol$ to $9 \zsol$, the $10^5 \kel$ WA maintains almost constant temperature. For sub-Solar metallicity, $\xi_5$ decreases steadily through the increase of abundance, but this trend is inverted for super-Solar metallicity. The $10^6 \kel$ WA, on the other hand, shows a rise in temperature $T_6$ throughout the metallicity variation, while $\xi_6$ remains constant within a factor of 1.4. 
\item[$\bullet$] There is an anti-correlation between $\Delta_5[\log(\xi/T)]$ and $\Delta_6[\log(\xi/T)]$; the $10^5 \kel$ WA shrinks in $\log(\xi/T)$ until it vanishes for $9 \zsol$, whereas the phase at $10^6 \kel$ progressively grows in $\log(\xi/T)$ from $0.4 \zsol$ to $9\zsol$. This trend is also depicted in the top and the middle right panels of \fig{multiphase} by the {\it open stars} and the {\it closed circles}. A $10^7 \kel$ phase starts appearing from $3 \zsol$, and $\Delta_7[\log(\xi/T)]$ increases with abundance. 
\item[$\bullet$] Interesting multiphase properties are exhibited by WA with varying abundance. $\Delta_{56}[\log(\xi/T)]$ is significant in the range $0.4 \le Z/\zsol \le 3$ for which $10^5 \,\,\rm{and}\,\, 10^6\kel$ gas are in pressure equilibrium with each other. Beyond this range for $Z/\zsol \ge 5$, the stability curves show the interesting feature of significant multiphase solution between the $10^6 \,\,\rm{and}\,\, 10^7\kel$ phases; a property depicted by only these three stability curves among all the parameter sets considered in this paper.
\item[$\bullet$] In the case of the parameter sets having super-Solar metallicity, we find an interesting phenomenon depicted by the stability curves. For $Z/\zsol \ge 3$, the WA stable phases ($5 \le \log T \le 7$) are seen to be in pressure equilibrium with gas at $\sim 10^4 \kel$, which might be representing the `broad emission line region (BELR) clouds'. Such a feature in the stability curve might indicate the connection between the X-ray and UV absorber.
\end{enumerate}

\section{Summary}
\labsecn{summary}

Absorption edges due to various partially ionized elements are seen in the soft
X-ray spectra of about half of the Seyfert1 galaxies and observed quasars.  
These absorption features are due to the presence of a
warm ($\sim \onlyten{5} - 10^{6.5} \kel$) gas along our line of sight to the AGN.
We have shown that the existence and nature of this absorption component
crucially depends on the shape of the ionizing continuum from the central source
which illuminates the gas, and also on the chemical composition of the absorbing
medium.  We have used the stability curve as a tool to study the
stability of the WA as a function of its various physical properties.

We have considered an ionizing continuum with two power-law components which
together reproduce the observed X-ray spectral index as well as the observed
values of $\alpha_{ox}$. Systematic investigation shows that if the X-ray
spectral index, $\alpha$, of the ionizing continuum intervening gas of
solar metallicity is flat with $\alpha \sim0.2$, then the AGN cannot have a WA. 
For
moderately steep X-ray spectra with $0.5 \lesssim \alpha \lesssim 1.1$, it is
possible to sustain a WA. A multiphase nature of the WA is, however, seen only
for $\alpha \sim 0.8$. It is an interesting coincidence that most of the 
observed quasars also have soft X-ray slopes similar to $\alpha \sim 0.8$ and 
in future studies we shall investigate whether such a coincidence leads to 
deeper physical implications. If the ionization continuum becomes even steeper 
then, instead of discrete phases, the absorbing gas exhibits a continuous 
distribution of $\xi/T$ and $T$. We have also verified that the nature of 
the WA is not affected by the higher energy cut-off of the ionizing continuum.

Temperature estimates for warm absorbing gas were generally accepted to be 
independent of the density of the gas, because the ionization and thermal 
equilibrium is believed to be achieved by the balance between photoionization and 
recombination. We show in this paper, in agreement with Rozanska \etal (2008), that the 
stability properties of highly dense WA might show density dependence influenced by 
sufficiently soft UV dominated ionizing continuum. This is a very interesting 
result because it implies that for such objects, we would not require independent 
estimates of density from observations; modeling the UV spectra itself can provide 
constraints on gas density.

The chemical composition of the absorbing medium plays a critical role in
determining its multiphase nature. The higher the metallicity of the medium, the
greater is the probability of a multiphase parameter space. The role of individual
elements or certain important groups of elements were also studied.  Amongst the
elements, iron, which is formed from type Ia supernovae no earlier than 1 Gyr, 
plays the most important role as a heating agent at higher 
temperatures ($\ge 10^5 \kel$) while oxygen is a significant cooling agent for 
$T \le 10^5 \kel$. The X-ray group (C, Ne, O, Fe), 
understandably, has significant influences, because the constituent elements 
are the ones which have important atomic transitions in the energy range relevant 
for warm absorbers ($0.3 - 1.5 \kev$). If the absorber
is over abundant in iron, oxygen or in the X-ray group of elements then the absorbing
medium is likely to be multiphase. A very interesting result seems to suggest 
itself from the study of influence of $\alpha$-elements (Ne, Mg, Si, S, Ar, Ca, Ti), 
which are the first metals formed in the Universe along with oxygen through the 
supernovae of type II. The thermal properties of the warm absorber change 
drastically from the time when the absorbing gas is essentially constituted of 
only $\alpha$-elements as metals to the time when the absorber is enriched enough 
to have the whole X-ray group included in its chemical composition. We have also 
observed that zero metal abundance gas does not produce WAs in AGN. Our analysis 
on the chemical composition of the absorbing gas shows that the warm absorber 
temperature range is very sensitive to the exact abundance of the absorber and there 
is an indication that the warm absorber might have super-Solar metallicity. While 
fitting soft X-ray data and determining the physical properties of the warm absorber 
in individual objects care should be taken to include the effect of super-Solar 
metallicities.

When we consider super-Solar metallicity, the stability curve shows a 
$10^4 \kel$ stable state is in pressure equilibrium with the WA stable
phases. This might imply some connection between the WAs and the BELR. 
We plan to take up this issue in future work.

We have used an upgraded version of {\tiny CLOUDY} (C07.02) where the most recent 
developments in atomic physics have been incorporated. This gives us the 
confidence to carry out a quantitative analysis on the multiphase nature of 
the WA presented in rigorous details in \secn{multiphase}. We have been able 
to provide specific trends of the behaviour of the WA as a function of 
the prevalent 
physical conditions like the shape of the ionizing continuum and the chemical 
composition of the gas. Comparison between observations of WA properties and 
simulations can be made more concrete by providing such quantitative measures. 
We will extend this exercise to greater details in our subsequent publications.

In this paper we have used simple broken power-laws to define the ionizing
continuum for the absorbing gas. However, the soft excess in Ultraviolet is
often modeled using blackbody components which peak at $T \sim 150
\ev$. Moreover the continuum spectra of AGN also has the disk blackbody
component at $20 \ev \lesssim T \lesssim 30 \ev$. We intend to investigate the
effects of such continua on the WA properties as the next step in this
systematic analysis.


\section*{acknowledgements}
SC would sincerely like to thank Ranjiv Misra and Raghunath Srianand for useful 
discussions and for suggestions which have helped the development of the paper. We 
would like to sincerely thank the anonymous referee whose comments have significantly 
improved the paper.  


\end{document}